\documentclass[10pt,journal,compsoc]{IEEEtran}

\usepackage{amsthm}
\usepackage{graphicx}
\usepackage{moreverb}
\usepackage[colorlinks=true,citecolor=red]{hyperref}
\usepackage{graphicx}
\usepackage{float}
\usepackage{amssymb}
\usepackage{amsfonts}
\usepackage{multicol}
\usepackage{amsmath}
\usepackage{mathtools}
\usepackage[mathscr]{euscript}
\usepackage{booktabs}
\usepackage{makecell}
\usepackage{multirow}
\usepackage{amsbsy} 
\usepackage[hypcap=false, font=small]{caption}
\usepackage[noend,ruled,linesnumbered]{algorithm2e}
\usepackage{cite}
\usepackage{xfrac}
\usepackage{bbm}
\usepackage{color, colortbl}
\definecolor{Gray}{gray}{0.9}
\usepackage[first=0,last=9]{lcg}

\usepackage[absolute,showboxes]{textpos}
\PassOptionsToPackage{draft}{hyperref}
\usepackage{subfig}

\begin{document}

\title{$S^2VC$: An SDN-based Framework for Maximizing QoE in SVC-Based HTTP Adaptive Streaming}

\author{\IEEEauthorblockN{Farzad Tashtarian$^\star$, Alireza Erfanian$^\star$, and Amir Varasteh$^\dagger$ \\
$^\star$ Department of Computer Engineering, Mashhad Branch, Islamic Azad University, Mashhad, Iran\\
Email: \{f.tashtarian, a.erfanian\}@mshdiau.ac.ir\\
$^\dagger$ Chair of Communication Networks, Department of Electrical and Computer Engineering\\
Technical University of Munich, Germany\\
Email: amir.varasteh@tum.de}}

\IEEEtitleabstractindextext{%
\begin{abstract}
HTTP adaptive streaming (HAS) is quickly becoming the dominant video delivery technique for adaptive streaming over the Internet. Still considered as its primary challenges are determining the optimal rate adaptation and  improving both the quality of experience (QoE) and QoE-fairness. Most of the proposed approaches have relied on local information to find a result. However, employing techniques that provide a comprehensive and central view of the network resources can lead to more gains in performance. By leveraging software defined networking (SDN), this paper proposes an SDN-based framework, named $S^2VC$, to maximize QoE metrics and QoE-fairness in SVC-based HTTP adaptive streaming. The proposed framework determines both the optimal adaptation and data paths for delivering the requested video files from HTTP-media servers to DASH clients. In fact, by utilizing an SDN controller and its complete view of the network, we introduce an SVC flow optimizer (SFO) application module to determine the optimal solution in a centralized and time slot fashion. In the current approach, we first formulate the problem as a mixed integer linear programming (MILP) optimization model. The MILP is designed in such a way that it applies defined policies, e.g. setting priorities for clients in obtaining video quality. Secondly, we show that this problem is NP-complete and propose an LP-relaxation model to enable $S^2VC$ framework for performing rate adaptation on a large-scale network. Finally, we conduct experiments by emulating the proposed framework in Mininet, with the usage of Floodlight as the SDN controller. In terms of improving QoE-fairness and QoE metrics, the effectiveness of the proposed framework  is validated by a comparison with different approaches.
\end{abstract}

}

\maketitle

\IEEEraisesectionheading{\section{Introduction}\label{sec:introduction}}
\IEEEPARstart{I}{n} recent years, the presence of the Internet and its applications in various aspects of our lives has exponentially increased. As mentioned in \cite{index2015cisco}, video streaming traffic, which has made up the largest portion of Internet traffic, will grows to be up to 75\% of the total Internet traffic by 2020. For many years, the UDP protocol has been employed to transfer multimedia traffic in the Internet. In recent years, much efforts has been spent to utilize TCP for multimedia transmission over the Internet. With HTTP, the employment of caches and also content delivery networks (CDNs) are possible, thus providing network scalability and traffic reduction. In addition to TCP's reliable transmission and cache-friendliness, streaming data with TCP allows for easy traverse of firewalls and NAT devices. As a result, deploying HTTP for multimedia transmission over the Internet has significantly risen. For instance, nowadays, more than 98\% of video traffic in cellular networks is transmitting via the HTTP protocol \cite{erman2011cache}.

\par 

The most widely used technique for TCP streaming is HTTP-based adaptive streaming (HAS). In this technique, a video file is divided into short duration segmented files, each of which is encoded at different bit rate levels and resolutions. Many companies have developed modifications of HAS systems, such as Smooth Streaming \cite{micsmooth}, HDS \cite{adobehttp}, and HLS \cite{applehttp}. In 2012, HAS was standardized by the Motion Picture Experts Group (MPEG) and named dynamic adaptive streaming over HTTP (DASH) \cite{stockhammer2011dynamic}. With DASH, a video is segmented into short segments encoded at various bit rates. This information is then stored in a media presentation description (MPD) file. \par 

There are several encoding methods for encoding a video. The two most widely used are advanced video coding (AVC) \cite{wiegand2003overview} and scalable video coding (SVC) \cite{schwarz2007overview}. In AVC-DASH, the client downloads the MPD file to obtain information from the server such as segment details, available bit rates, etc. Then, the appropriate bit rate is selected and HTTP streams these segments to the end-users. Multiple copies of a video are encoded with different bit rates and stored on the media server, thus leading to storage overhead. In contrast SVC is a layered video codec in which the video stream is encoded in a base layer and in one or more enhancement layers. The base layer provides the minimum usable quality (i.e. resolution and frame rate) for the clients. Better quality alternatives are available in the enhancement layers. SVC has several advantages over AVC, including higher web caching performance, lower bandwidth usage, and higher quality of experience (QoE) for the users \cite{sanchez2011idash}. In addition, as shown in \cite{sanchez2011idash}, SVC-DASH is able to provide media streaming to a larger number of users in environments with heterogeneous devices. Moreover, although SVC-DASH requires less buffer size, it increases storage-efficiency and QoE \cite{huysegems2012svc}.

In general, maximizing QoE and improving QoE-fairness can be achieved by employing an optimal adaptation in SVC-based streaming. This can be performed by three approaches: 1)\textit{ Purely client-based}, 2)\textit{ Client-based assisted by network elements}, and 3)\textit{ Network-based}. In purely client-based, according to the local parameters (e.g., available bandwidth and buffer occupancy), the client adapts the quality of the video by adding or removing one or more video enhancement layers. However, since clients are unaware of the whole network topology and its current state, this adaptation technique is sub-optimal and can be disadvantageous in shared network environments \cite{seufert2015survey}. For instance, based on the network status, some clients may change the video quality frequently (i.e., bit rate oscillation), which decreases the user QoE. In the case of client-based assisted by network elements, client adaptation can be assisted by an element in the network, such as proxy servers. In network-based, by utilizing the complete view of the current network state, rate adaptation can be performed by a centralized controller \cite{thomas2015enhancing}. Upon consideration of these methods, it is obvious that the network-based approach is more advantageous from the user QoE and QoE-fairness points of view. \par

Software defined networking (SDN) has recently emerged as a networking paradigm. In this architecture, the data and control plane are decoupled, which sheds new light on networking technology \cite{mckeown2008openflow}, \cite{mckeown2009software}. The data plane consists of hardware or software elements that are dedicated to forwarding flows and packets. In Contrast, with an SDN controller, the control plane manages the data plane elements. By applying the current study's innovations in the network, we can develop network applications to the control plane that are able to communicate with the SDN controller through application programming interfaces (APIs), e.g, RESTful API. In fact, SDN provides a flexible platform which can perform adaptive routing algorithms for different network applications. In addition, SDN can modify traffic to increase QoE for certain traffic flows (e.g., multimedia streaming traffic). \par
In the present work, we propose  $S^2VC$ as an SDN-based framework to maximize the QoE and QoE-fairness of SVC-based HTTP adaptive streaming. In this framework, a SVC flow optimizer (SFO) application module is introduced which is run by a controller to centrally adjust the clients' adaptation rate. In fact, by employing a holistic network view provided by the SDN controller and collecting some critical information from HTTP-media servers and clients, the SFO jointly determines the optimal quality adaptation and flow paths for client requests. In summary, the present study's  main contributions can be described as below:
\begin{itemize}
  \item Describe SDN-based framework for SVC-based HTTP adaptive streaming
  \item Consider substantial QoE metrics and fairness in the process of designing architecture of the proposed framework
  \item Propose a mixed integer linear programming (MILP) model to jointly determine the optimal data paths for delivering the requested video files and the quality adaptation
  \item Present a linear programming (LP) relaxation model of the proposed MILP model
  \item Implement the proposed framework and evaluate its performance in comprehensive scenarios
\end{itemize}

The remainder of this paper is organized as follows. In Section 2, related work is presented. Section 3 introduces and elaborates on the proposed $S^2VC$ framework and its details. The performance evaluation is presented in Section 4. Finally, Section 5 concludes the paper.

\section{Related Work}
As mentioned before, in current DASH-based approaches, clients implement adaptation techniques locally. This allows clients to select their preferred video quality without being informed about the whole network view. Recently, Seufert \textit{et al.} in \cite{seufert2015survey} demonstrated that the existence of a controller or a client-proxy connection can improve the client video quality and provide fairness in shared resource usage. In this section, we discuss the existing three main solutions for determining the optimal adaptation. \textit{purely client-based}, \textit{client-based assisted by a network element}, and \textit{network-based} adaptation techniques. In the \textit{purely client-based} technique, a client performs quality adaptation based on its local parameters, such as network throughput, occupied buffer size, etc. In the second approach, these clients can be assisted by the information provided by network elements, such as proxy servers. On the other hand, \textit{network-based} methods use a central network element to perform rate adaptation on behalf of the clients.

\subsection{Purely Client-Based Adaptation}
As shown in \cite{akhshabi2011experimental,huang2012confused} DASH video players in clients are responsible for quality adaptation in order to optimize the QoE objectives, e.g., initial buffering time minimization, stalling minimization, and quality maximization. In \cite{jiang2012improving,li2014probe}, the authors propose a general bit rate adaptation
framework that consists of a set of methods
striving to achieve a trade-off between video stability,
fairness, and efficiency. A stateful bit rate selection heuristic algorithm is used to achieve a biased interaction between the bit rate and estimated bandwidth. However,  this client-based quality adaptation leads to uneven bandwidth competition, which intensifies when a large number of clients use the shared network resources \cite{bentaleb2016sdndash}. Moreover, resource fairness does not reduce QoE-fairness, specially in heterogeneous environment \cite{bentaleb2017sdnhas}. This unfair resource usage, coupled with unpredicted network traffic bursts, can cause frequent changes in the quality layers of clients (client quality oscillations), which significantly reduces QoE. Therefore, an optimization model should be developed to trade-off between different QoE parameters, such as the video quality maximization and minimization of bit rate changes, while also meeting network constraints. \par 
In order to achieve this goal, various algorithms have been developed through the consideration of different parameters such as estimated network bandwidth, network throughput, and received network feedback signals (e.g., congestion occurrence) \cite{yin2015control,sieber2013implementation,schierl2011priority,xiang2012adaptive}. Sieber \textit{et al.} in \cite{sieber2013implementation} present an algorithm to achieve maximum QoE by reducing the quality switching frequency. The authors of \cite{xiang2012adaptive} employ Markov Decision Process to determine an optimal streaming strategy. Their method is developed in a wireless network environment and aims to optimize the QoE in terms of video playback interruption, average playback quality, and playback smoothness. Additionally, \cite{schierl2011priority} provides a technique to overcome network fluctuations, which is typical in mobile networks. These authors first prioritize the SVC layers. Then, the base layer (which has the most priority) is delivered to user via RTP flows. Thereafter, additional enhancement layers are delivered based on their priority. Further, \cite{famaey2013merits} implements and compares different heuristic algorithms based on AVC and SVC. It is shown that AVC performs better than SVC in situations with high delay while, in networks with unpredictable interruptions, SVC has superior performance. Moreover, Lee \textit{et al.} in \cite{lee2014caching} investigate the impact of cache storage and report that it can increase oscillation. They then propose a method to solve this. However, it is clear that, by using purely client-side sub-optimal adaptation decisions and also with the absence of a central controller for clients, the video quality changes frequently and  QoE reduction ultimately occurs \cite{akhshabi2011experimental,kuschnig2010evaluation}. Furthermore, with a central controller, it is possible to enforce different management policies, such as various subscription policies (e.g., Gold, Silver, Bronze) \cite{krishnamoorthi2013helping}. The following discussion explores quality adaption techniques assisted by a network element. \par
\subsection{Client-Based Adaptation Assisted by Network Elements}
As we mentioned before, specific network elements can assist clients for client-based rate adaptation. Using the information provided by network elements, a client can enhance its quality adaptation procedure \cite{thomas2015enhancing}. Petrangeli \textit{et al.} in \cite{petrangeli2016qoe} utilizing an SDN controller with a holistic network view, assist client-side quality adaptation in an AVC-DASH streaming method. Additionally, in \cite{georgopoulos2013towards}, the authors propose an SDN-based video streaming approach to fairly maximize the QoE of multiple competing clients in a shared network environment. Likewise, the proposed strategy in \cite{kleinrouweler2016delivering} is based on SDN. In fact, the authors of \cite{kleinrouweler2016delivering} employ two main approaches to optimize QoE in terms of number of quality changes and fairness. In the first approach, the SDN controller determines the video quality of the clients. In the second one, a queue for each client is developed to provide dynamic rate adaptations. Although these studies utilize some network elements to improve QoE, the determination of optimal adaptations is performed separately by clients. Thus, these strategies do not lead to efficient share usage of network resources. Bentaleb et al. in \cite{bentaleb2016sdndash} address HAS scalability issues, including video instability, QoE-unfairness, and network resource under-utilization. To cope with these issues, they maximize the QoE per client.
Moreover, \cite{bentaleb2017sdnhas} mitigates the main drawbacks of SDNDASH\cite{bentaleb2016sdndash}: scalability , communication overhead, and the support of client heterogeneity.\par

There are some works however, that have mainly focused on QoS-aware video traffic routing \cite{egilmez2011scalable,egilmez2013optimization,egilmez2014distributed,zhu2013design,egilmez2012openqos,li2012efficient}. In fact, these studies have investigated the QoS-aware video flow routing in OpenFlow/SDN-enabled networks. Indeed, the base quality layer is considered as the flow with the maximum QoS priority, while other layers are transferred in the best effort manner \cite{egilmez2011scalable,egilmez2013optimization}.  Thereafter, by considering the QoS priorities; the routing algorithm finds the shortest path between the media server and client. The authors extend their previous work in \cite{egilmez2012openqos}, in which they consider end-to-end QoS in multi-domain SDN networks. In addition, Egilmez et al. \cite{egilmez2014distributed} design a QoS-aware controller to deliver multimedia flows in OpenFlow-enabled networks. In this approach, they classify input traffic based on data flow types. Then, multimedia flows are routed through QoS-guaranteed paths and other flows are delivered to the destination via the shortest paths. These studies do not discuss quality adaptation techniques. In addition, the authors of \cite{li2012efficient} provide a model to determine the $k$ shortest paths and to then assign these to each $k$ SVC layer. They extend their research in \cite{zhu2013design}, in which they considered additional parameters, such as bandwidth and delay, and also enhance the model by using the max flow algorithm. Although these techniques can improve client-side decisions, these decisions are still sub-optimal as there is no independent central element for decision making. Furthermore, clients might not be able to perform the decisions made by the controller \cite{kleinrouweler2016delivering}. Therefore, central-based quality adaptation can be suggested.
\subsection{Network-Based Adaptation}
Several works use a proxy/controller as a central rate adaptation decision-making authority \cite{bouten2012qoe,mok2012qdash,el2015qoe}. In \cite{bouten2012qoe}, the authors propose an approach that actually employs a proxy server to monitor video quality requests. In this approach, a proxy server periodically solves an optimization problem in order to determine the maximum segment quality levels which the clients can download. This is achieved based on the current network status and a specific objective function. In fact, whenever a client-based adaptation reduces overall fairness, the proxy server is able to replace its video quality decision with that of the client. Therefore, a certain level of QoE can be guaranteed among some or all clients. In addition, \cite{hsiao2010design} suggest utilizing an intermediate node to perform the quality adaptation. Actually, this intermediate node acts as a gateway for clients. Practically, it uses the clients' bandwidth estimation to determine the optimized video quality of SVC clients. Moreover, Mok \textit{et al.} \cite{mok2012qdash} design a proxy architecture to prevent frequent and drastic oscillations and to allow a specific amount of change in video quality in each step. Unlike ours, this approach is not scalable, since all the network traffic to the video server must pass through a proxy server. Furthermore, \cite{el2015qoe} presents reactive and proactive QoE optimization approaches. In the reactive method, client-based quality adaptation is achieved using network information. By considering client buffer utilization and quality adaptation in the controller, the second method provides a higher and fairer video quality to clients. \par

Only a few works \cite{yue2015joint,uzakgider2015learning}, however, have focused on jointly determining routing and providing quality adaptation. \cite{yue2015joint} proposes a method based on Markov Chains, which determines the quality adaptation in the controller and selects the optimal video quality of the clients. Their approach first determines the $N$ shortest-path routes and then, based on priorities, sends the video layer using one of the paths. In contrast, the approach in \cite{uzakgider2015learning} utilizes the optimal paths instead of the shortest ones. However, in \cite{yue2015joint} and \cite{uzakgider2015learning}, the authors only determine an optimal data path and do not consider fairness in their proposed solution. By running the algorithm for each incoming client (or request), they find a solution regarding the available resources. Moreover, \cite{yue2015joint} and \cite{uzakgider2015learning} do not target DASH video streaming since they focuse on the UDP protocol. Cetinkaya \textit{et al.} \cite{cetinkaya2014sdn} develop an optimization model to maximize video quality by the selection of optimal paths for different SVC layers over SDN. Finally, it informs clients about the selected bit rate to be applied by them. In addition, in \cite{civanlar2010qos}, the authors propose an approach which delivers the base layer to clients via a lossless path and uses other paths to route the remaining layers.

\section{The proposed $S^2VC$ framework}
\label{sec3}
In this Section, by leveraging the SDN conceptual model, the current study addresses the issues of client quality adaptation, \textit{quality of experience} (QoE) and QoE-fairness. Before introducing the details of $S^2VC$, we shall describe the considered QoE and the QoE-fairness metrics. \par

\subsection {The defined QoE metrics and QoE-fairness}
As mentioned earlier, a DASH player first downloads the media presentation description (MPD) file that provides the segment information and then requests each segment layer individually. However, we assume that the DASH players can determine and send the maximum supported video quality, according to their resource capacity. Nevertheless, for each client, these values can be adjusted by $S^2VC$. In the proposed framework, the time for buffering a segment is considered fixed, even though it is possible that a DASH player suggests a deadline in which all requested layers of a segment must be downloaded. \par
To achieve better video quality, layers with higher data rates should be buffered by the DASH client. Many functions have been proposed to measure the quality of the received video files  \cite{schwarz2007overview} as it has been shown that receiving layers with higher bit-rates can promote the quality of a video. Therefore, in this study, without loss of generality and for ease of explanation, the number of continuous buffered layers by a DASH client is considered as the achieved video quality. $S^2VC$ considers the following QoE metrics and QoE-fairness:
\begin{itemize}
\item \textit{Start-up delay}: The time period between sending a video request and starting to render the first frame.
\item \textit{Number of stalls}: The stall phenomenon occurs when the buffer space of a client reaches its minimum threshold value (possibly be zero) during the playback of the requested video. At this time, the DASH player stops rendering and waits to download the current segment. 
\item 	\textit{Average video quality}: The average video quality of all received segments. In other words, the average number of buffered layers of each segment of a video file. 
\item \textit{Average number of video quality switches}: This metric is expressed as the average number of video quality switches between any successively received segments. 
\item \textit{Average intensity of video quality switches}: In addition to the average number of video quality switches, we take the average intensity of video quality oscillations into account. This metric indicates how much the quality of any two successive segments oscillates. In other words, this can be defined as the average difference between the number of downloaded layers of any successively received segments. 
\item \textit{QoE-fairness}: The variance of video quality received by clients at the same time.
\end{itemize}\par

\begin{figure}[t]
\centering
\includegraphics[width=\linewidth]{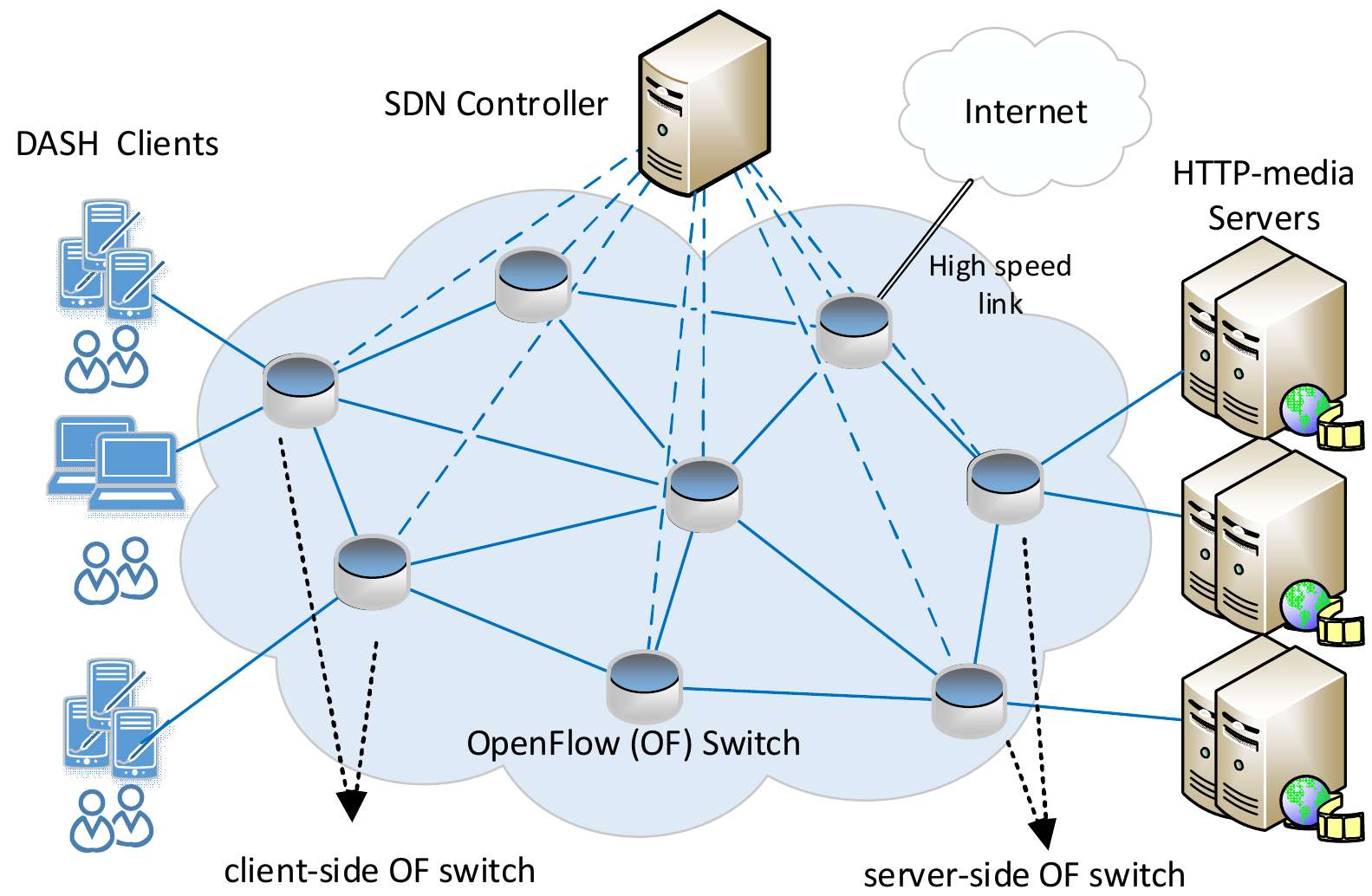}
\caption{\textbf{The communication topology of $S^2VC$ framework }}
\label{fig1}
\end{figure}

\subsection {The Details of $S^2VC$}
The schematic communication topology of the proposed framework is illustrated in Fig. \ref{fig1}.  In this architecture, by leveraging the SDN paradigm, we introduce an SVC flow optimizer (SFO) application module as the core part of $S^2VC$. In fact, the SFO is employed by the SDN controller to efficiently enable an SVC-based HTTP adaptive streaming service for responding to the requests of DASH clients. This communication model can be utilized in the network edge and the requests of DASH clients will be served by HTTP-media cache servers if the requests hit; otherwise they must be forwarded to the origin server located on the Internet. However, the present study assumes that all DASH client requests are served by the local HTTP-media servers. 
\begin{figure}[t]
  \centering
\includegraphics[width=.9\linewidth]{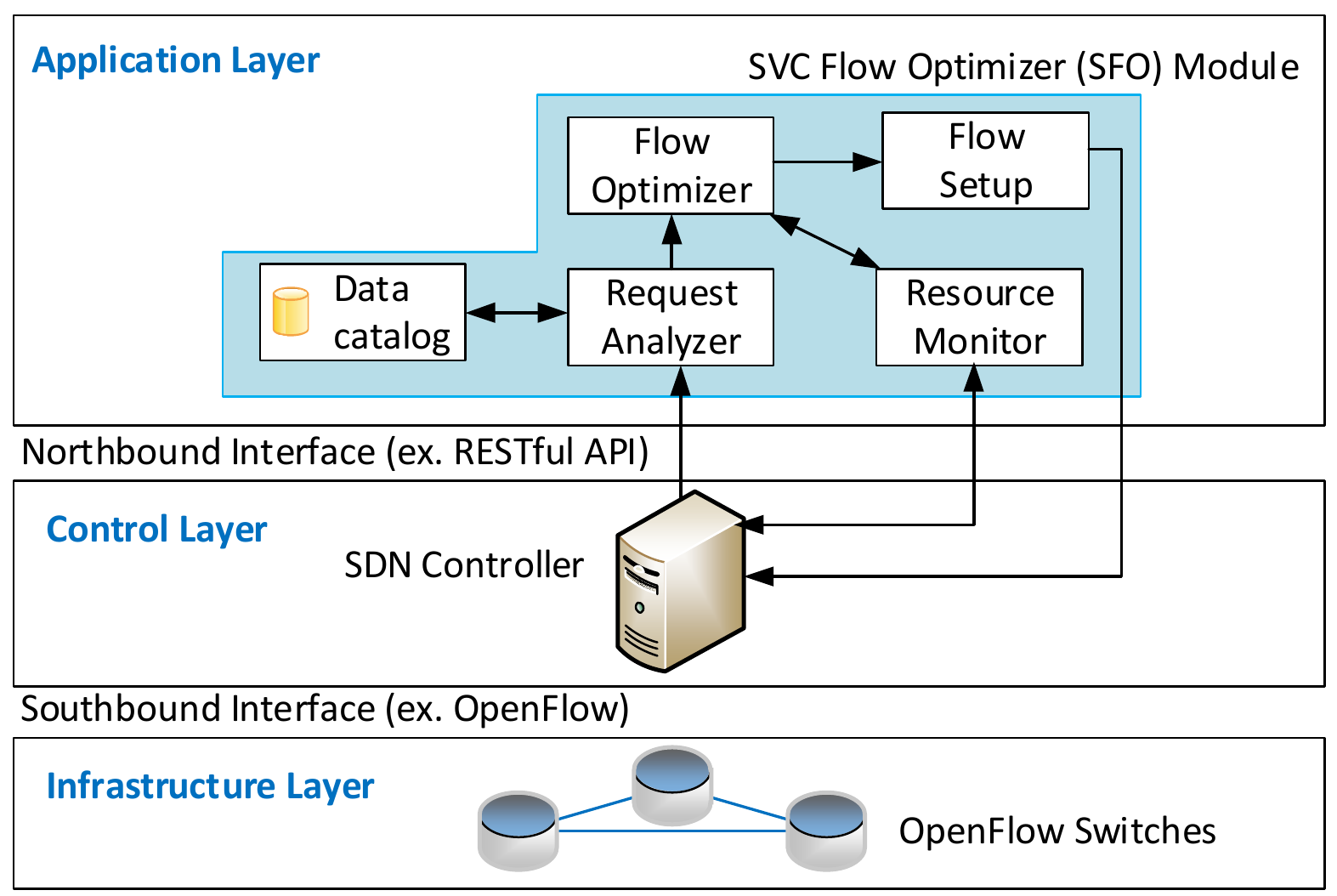}
            \caption{\textbf{$S^2VC$ three layered architecture model }}
            \label{fig7}
\end{figure}

To clarify and investigate the operation of the proposed approach, we also illustrate the three layer architecture model of $S^2VC$ which consists of the \textit{infrastructure, control}, and \textit{application} layer (see Fig. \ref{fig7}). As illustrated, OpenFlow (OF) switches, the SDN controller, and the SFO module are placed in the infrastructure, control, and application layer, respectively. \par
\begin{figure}[t]
\centering
\includegraphics[width=\linewidth]{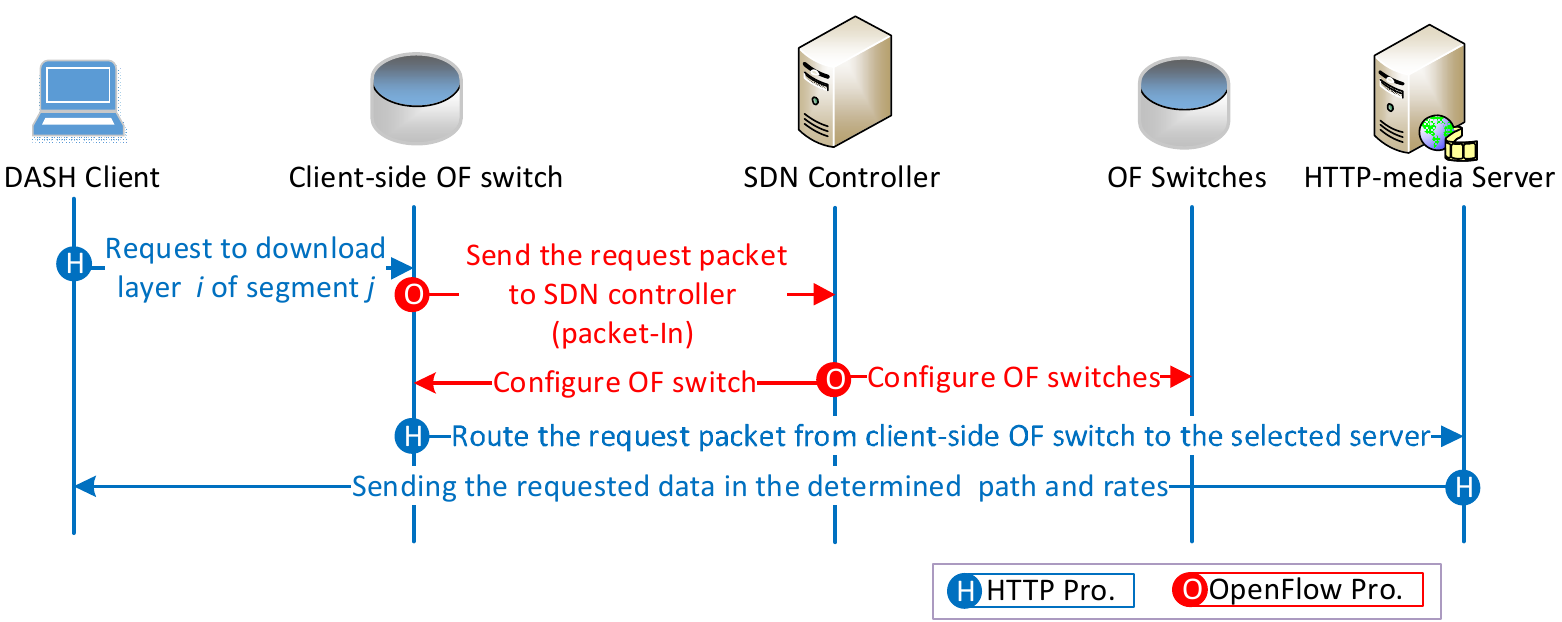}
\caption{\textbf{Message exchanges for download layer $i$ of segment $j$ by a DASH client}}
\label{fig6}
\end{figure}

$S^2VC$ operates in a time slotted manner, in which, in each time slot $\tau$ SFO must be executed to determine an optimal solution. Both DASH clients and HTTP-media servers are connected to the SFO via OF switches and the SDN controller. DASH client request must be sent to the SDN controller for determining the optimal solution through SFO. Thus, to have the minimum overhead of OF switches, we configure that DASH clients connect to a \textit{default HTTP-media server} with a specific IP and port address. Therefore, the client-side OF switch (the first hop OF switch of the DASH client) receiving a packet with destination IP and port address of the default HTTP-media server, is configured to be forwarded it to the SDN controller as a Packet-In.
Considering the received requests and available resources, e.g. link bandwidth, SFO determines an optimal solution for the addressed problem. Thereafter, through the SDN controller, SFO configures the selected HTTP-media servers and OF switches in order to launch the data transmission process, which is achieved by sending a Packet-Out message to the OF switch. The message exchanges among these items are depicted in Fig. \ref{fig6}. The SFO is comprised of five components: the request analyzer, resource monitor, flow optimizer, flow setup, and data catalog component. 

\begin{figure}[b]
\centering
\includegraphics[width=\linewidth]{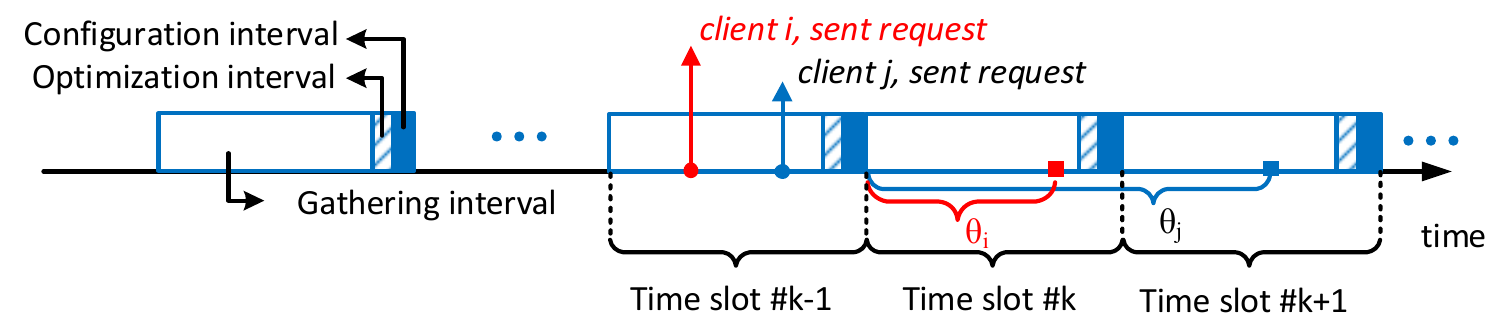}
\caption{\textbf{The time slot in $S^2VC$}}
\label{fig4}
\end{figure}

\textbf{Request Analyzer Component (RAC)}:
After receiving the Packet-In, the SDN controller forwards it to the RAC. RAC is responsible for receiving the DASH client requests at each time slot $\tau$, at which point it launches the flow optimizer component. In fact, each time slot $\tau$ is divided into three unequal intervals: the gathering, optimization, and configuration interval (see Fig. \ref{fig4}). The two short optimization and configuration intervals are allocated for running the flow optimizer component and configuring OF switches to start data transmission, respectively. According to the output of the flow optimizer component, it is possible that data transmission takes longer than the time slot. In the gathering interval, RAC buffers the received requests from the DASH clients and performs some pre-processing on the collected requests to prepare them as an input parameter for the flow optimizer component in the next time slot.\par

In the pre-processing step, RAC must extract some key values from the client requests. According to the received requests in each gathering interval,  RAC creates two sets, namely $\mathbb{C}$ and $\mathbb{D}$. The first set, $\mathbb{C}$, includes of the client-side OF switches with their connected DASH clients, where $\mathcal{N}_i$ indicates the DASH clients connected to the client-side OF switch $i$ and $\mathbb{D}$ is the set of DASH clients that request to download desired layers of demanded segments. Since the flow optimizer component needs to take all gathered requests into account, RAC assigns a unique identification (ID) for each \textcolor{black}{request by parsing the headers of transport and the application layers of the incoming packets, e.g. a combination of incoming sockets and the file name of the requested segment.} Furthermore, RAC suggests $\theta_{c}$ as the deadline for buffering the requested segment (the waiting time of DASH client $c\in \mathbb{D}$ in the gathering interval can be taken into account) and $m_c$ as the maximum video quality layer of the segment determined by client $c\in \mathbb{D}$. We note here that the values of  $\theta_{c}$ can be assumed as fixed. However, to achieve a more flexible model, it is possible these be determined by DASH players or RAC as appropriate. As illustrated in Fig. \ref{fig4}, the two clients, $i$ and $j$, send their requests in time slot $k-1$. Both clients must wait for the optimization interval of the next time slot (time slot $k$) to obtain an optimal solution. As shown, the deadlines for the requested video files in time slot $k-1$ are considered as being from the beginning of the next time slot or time slot $k$.\par


\textbf{Resource Monitor Component (RMC)}: Before describing RMC, let us define graph $G=\{V,E\}$, where $V$  is the set of \textcolor{black}{ OF switches, DASH clients, and HTTP-media servers; and set $E$ represents the edges of $G$, where ${e_{ij}=1}$ if a direct communication link exists between the two entities of $i$ and $j \in V$. At the beginning of each time slot, RMC uses RESTful APIs to obtain the available links' bandwidth from the SDN controller. The measured values are stored in a two-dimensional array $B$, in which $b_{ij}$ indicates the available bandwidth between $i$ and $j$ $\in V$. The connectivity among OF switches, DASH clients, and HTTP-media servers can easily be inferred from $B$. In fact, this component constructs graph $G=\{V,E\}$ and the available bandwidth among the nodes in $V$. }\par

\textbf{Data Catalog Component}:
This component provides critical meta-data \textcolor{black}{about the available video files in HTTP-media servers. The meta-data includes the list of servers, stored file segments, layers, and their properties, such as: the size and bit-rate of each layer.}
We define $\mathbb{S}$ as the set of \textcolor{black}{HTTP-media servers, in which at least one requested layer $l$ by DASH client $c\in\mathbb{D}$ is stored, where $l\leq m_c$. In fact, the data catalog component sets $a^s_{cl}=1$ if layer $l$ requested by $c$ is available in server $s$, else  $a^s_{cl}=0$. The data catalog component also uses $\delta_{cl}$ to show the size of layer $l$ requested by DASH client $c\in \mathbb{D}$.} 

\begin{table}[t]
    \centering
    \small
    \caption {\textbf{Notations}}
    \label{tab:table1}
    \begin{tabular*}{1\linewidth}{c|l}
    Notation & \hspace{2.5cm} Description \\ \hline
     $V$ & The set of OF switches, DASH clients, and \\~&HTTP-media servers\\
     $E,e_{ij}$ & The two-dimensional binary array $E$ represents \\~&edges among $V$, where ${e_{ij}=1}$ indicates a direct\\~& communication link exists between two entities of\\~& $i$ and $j \in V$\\
     $\mathbb{D} $ & The set of DASH clients\\
     $\mathbb{C}$ & The set of client-side OF switches\\
     $\mathbb{S}$ & The set of HTTP-media servers\\
     $\mathcal{N}_i$ & The set of DASH clients connected to client-side\\~& OF switch $i$ \\
     $\tau$ & Time slot duration\\
     $\theta_c$ & The offered deadline for delivering the requested\\~& segment from  HTTP-media server to client $c\in \mathbb{D}$ \\
     $m_c$ & The desirable maximum video quality determined\\~& by client $c\in \mathbb{D}$\\
     $B,b_{ij} $ & B is a two-dimensional array where $b_{ij}$ shows the\\~& available bandwidth between $i$ and $j\in V$ \\
     $A,a^s_{cl} $ & A is a three-dimensional binary array where\\~& $a^s_{cl}=1$ represents that layer $l$ requested by $c\in \mathbb{D}$\\~& is reachable through $s\in \mathbb{S}$ \\
     
     $t^{cl}_{ij} $ & The optimal data rate for transmitting the\\~& requested layer $l$  by DASH client $c$ that must be\\~& delivered from $i$ to $j\in V$ \\
     $\delta_{cl}$ & The size of layer $l$ requested by DASH client $c$ \\
     $\bar{\delta}_{c} $ & The average layer size of requested segment by\\~& client $c$ \\
     $\omega^s_{cl}$ & The binary variable determines that whether the\\~& requested layer $l$ by client $c$ must be served by\\~& server $s$ ($\omega^s_{cl}=1$) or not\\
      
     $T_{c}$ & The average video quality of the client $c$ from the\\~& beginning of its work until now\\ 
     $I_{c}$ & The average intensity of video quality switches\\~& for client $c$\\
     $N_{c}$ & The average number of video quality switches \\~&for client $c$\\
     \hline
    \end{tabular*}
\end{table}
\textbf{Flow Optimizer Component (FOC)}: In this component, we introduce a Mixed Integer Linear Programming (MILP) model to jointly determine optimal quality adaptation and data paths for delivering the requested data from HTTP-media servers to DASH clients, while maximizing the fairness and QoE. To find an optimal solution, a number of constraints must be satisfied. Before describing these, we define the input parameters as follows: $\mathbb{C},\mathbb{D}, \text{ }\theta_{c} \text{, and } {m_c}$ provided by RAC; $B \text{ and graph } G$ prepared by RMC; and $\mathbb{S} \text{ and } {A}$ obtained from the data catalog component. Table \ref{tab:table1} provides the main notations.\\
Let binary variable $\omega^s_{cl}$ determine whether server $s$ serves the requested layer $l$ by client $c$ ($\omega^s_{cl}=1$) or not ($\omega^s_{cl}=0$).
Thus, the following constraint states that one server can serve the requested layer:

\begin{equation}
   \sum_{s\in \mathbb{S}} \omega^s_{cl} \leq 1, \forall c \in\mathbb{D},l=1:m_c \label{eq2}
\end{equation}
The next constraint states that if server $s$ has a copy of requested layer $l$, then it can be selected as a candidate to respond to that client:
\begin{align}
    0\leq \omega^s_{cl}\leq a^s_{cl}, \forall s\in \mathbb{S}, \text{ } c \in\mathbb{D},l=1:m_c \label{eq11}
\end{align}
The third constraint guarantees that each client $c\in \mathbb{D}$ buffers a valid sequence of layers. In other words, if client $c$ receives layer $l\leq m_c$ from any server, then all layers from 1 to $l-1$ must be transmitted to it. Thus, we have:
\begin{align}
   \sum_{s\in\mathbb{S}}(\omega^s_{c(l+1)}-\omega^s_{cl})\leq0,  \forall  c \in\mathbb{D},l=1:m_c-1  \label{eq1}
\end{align}

Let $t^{cl}_{ij}$ be a data rate at which $i \in V$ sends the $l^{th}$ layer requested by  $c \in \mathbb{D}$ to $j \in V$ during $\theta_c$ units of time. To determine the data rates for transmitting all the requested layers from the optimal selected \textcolor{black}{HTTP-media servers in $\mathbb{S}$} to the DASH clients in  $\mathbb{D}$, the following constraints must be satisfied:
\begin{align}
   \theta_c(& \sum_{j\in V} e_{ij}t^{cl}_{ij}- \sum_{j\in V} e_{ij}t^{cl}_{ji}) \label{eq3}=\nonumber\\& \left\{\begin{array}{ll}\delta_{cl}\omega^i_{cl} \hspace{1.72cm},\forall \text{ }i \in \mathbb{S}, \text{ } c\in\mathbb{D}, \text{ }l=1:m_c \\0\hspace{.98cm},\forall \text{ }i \in V-\{\mathbb{S},\mathbb{C}\}, \text{ } c\in\mathbb{D},\text{ }l=1:m_c\\-\delta_{cl}\sum_{s\in \mathbb{S}}\omega^s_{cl}\hspace{.3cm},\forall \text{ } i\in\mathbb{C},\text{ }c\in\mathcal{N}_i,\text{ }l=1:m_c\\\end{array}\right. 
\end{align}
\textcolor{black}{Constraint (\ref{eq3}) covers three different cases: if $i$ is an HTTP-media server  ($i\in \mathbb{S}$), the constraint (\ref{eq3}) forces server $i$ to generate $\frac{1}{\theta_{c}}\delta_{cl}\omega^i_{cl}$ amount of traffic, where $\theta_{c}$ is the specified deadline by RAC for buffering each requested layer $l$ with size $\delta_{cl}$. In the second case, if $i$ is an OF switch, then $\forall \text{ } c\in\mathbb{D} \text{ and }l=1:m_c$, the total incoming traffic must be equal to its total outgoing traffic. Finally, the last case states that the all data of layer $l$ transmitting from server $s$ to client $c$ must be received by client-side OF switch $i$ if $(c\in\mathcal{N}_i)$. The following bandwidth constraint specifies an upper bound for the generated traffic on each link:}
\begin{align}
\sum_{c\in \mathbb{D}}\sum_{l=1:m_c} e_{ij}t^{cl}_{ij}\leq b_{ij}, \forall \text{ } i,j \in V\label{eq10}
\end{align}
 To consider the QoE-fairness and QoE metrics in our proposed MILP model, a number of constraints should be satisfied. First, we shall focus on the QoE-fairness in serving DASH clients in each time slot. Note, in this study, \textcolor{black}{ the quality of the received segment is measured based on the number of its received layers. This index will be compared to other well-known quality measuring methods in the performance evaluation. Hence, to achieve a fair video quality among the various maximum supported video qualities of the clients, we present the following constraint:
\begin{align}
 m_c-\sum_{s\in\mathbb{S}}\sum_{l=1:m_c}\omega^s_{cl}\leq Q m_c,  \forall c \in \mathbb{D}\label{eq12}
\end{align}}
In fact, Eq. (\ref{eq12}) determines the biggest gap, denoted by $0\leq Q\leq 1$, between the supported maximum layer $m_c$ and the maximum of layer served by the network for each client $c$. It is obvious that, by decreasing $Q$ in the objective function, the variance of the video quality received by clients will reduce and consequently the QoE-fairness will increase. The impact of $Q$ on QoE-fairness is comprehensively investigated in the performance evaluation. \par
We note that the first two QoE metrics $\textit{ start up delay }$ and $\textit{number of stalls}$, are taken into account by selecting the appropriate value for $\theta_c$. In fact, $\theta_{c}$ forces the MILP model to forward the requested layer by determining the optimal data rates (see Eq. (\ref{eq3})). However, for the other QoE metrics, namely $\text{ }\textit{average video quality}$, $\text{ }\textit{number} \text{ and  }\textit{intensity video quality switches}$, we need to keep track of their treatments in prior time slots. Let $\bar{\lambda}_c$ be the total video quality of the segments downloaded by the client $c$ from the time its first request is sent until the current time slot. Note that the number of downloaded layers for each segment is assumed to be the received video quality of that segment. Thus, the normalized average video quality of client $c$ from the beginning of its work until now, denoted by $T_c$, is obtained through the following constraint: \textcolor{black}{
\begin{align}
\frac{1}{\varphi_c}(\bar{\lambda}_c+\sum_{s\in\mathbb{S}}\sum_{l=1:m_c}\omega^s_{cl})=T_c  T_{max}, \forall c \in \mathbb{D}\label{eq13}
\end{align}}
where $T_{max}=max\{\bar{\lambda}_i+m_i\mid\forall i \in \mathbb{D}\}$ and $\varphi_c$ and \textcolor{black}{ $\sum_{s\in\mathbb{S}}\sum_{l=1:m_c}\omega^s_{cl}$ }are the total number of requested segments and the achieved layers of the video requested in the current time slot by client $c$, respectively. 

Let $\bar{\mu}_c$ and $\bar{\nu}_c$ be the total intensity and the total number of video quality switches by DASH client $c$ while obtaining segments from the beginning of its operation until the current time slot, respectively. Hence, the normalized average intensity of video quality switches, $I_c$, for client $c$ can be obtained as follows:
\textcolor{black}{
\begin{align}
   \frac{1}{\varphi_c}(\bar{\mu}_c+\mid\sum_{s\in\mathbb{S}}\sum_{l=1:m_c}\omega^s_{cl}-\bar{l}_c\mid) \leq I_c  I_{max}, \forall c\in \mathbb{D}\label{eq116}
\end{align}
where $I_{max}=max\{\bar{\mu}_i+m_i\mid\forall i \in \mathbb{D}\}$. In Eq.(\ref{eq116}), $\mid\sum_{s\in\mathbb{S}}\sum_{l=1:m_c}\omega^s_{cl}-\bar{l}_c\mid$ shows the difference in video quality among the layers received in the previous time slot denoted by $\bar{l}_c$, and the achieved  layers in the current time slot, as well as the notation $\mid.\mid$ indicating the absolute operation. Now, if $\mid\sum_{s\in\mathbb{S}}\sum_{l=1:m_c}\omega^s_{cl}-\bar{l}_c\mid>0$, we can conclude that the quality of the segment received in this time slot oscillates with respect to the perceived quality in the previous time slot. This statement can be formulated as follows: 
\begin{align}
  \mid \sum_{s\in\mathbb{S}}\sum_{l=1:m_c}\omega^s_{cl}-\bar{l}_c\mid\leq\nu_cm_c,\forall c\in \mathbb{D}\label{eq17}
\end{align}}
where the defined binary variable  $\nu_c$ determines whether the quality oscillation of the received segment occurs or not. Therefore, if $\nu_c=1$, the quality of the video oscillates else it remains unchanged. Thus, the normalized average number of video quality switches $N_c$ for client $c$ is obtained as follows:
\begin{equation}
   \frac{1}{\varphi_c}(\bar{\nu}_c+\nu_c)\leq N_c  N_{max},\forall c\in \mathbb{D}\label{eq107}
\end{equation}
where $N_{max}=max\{\bar{\nu}_i+1\mid\forall i \in \mathbb{D}\}$. Therefore, the MILP used in FOC can be represented as:
\begin{align}
 &\textit{\textbf{minimize}}\hspace{.2cm}  \alpha Q+\frac{1}{len(\mathbb{D})}\sum_{c\in\mathbb{D}}(\beta_{1c}I_c+\beta_{2c}N_c-\beta_{3c}T_c)\nonumber\\&\hspace{4.5cm}+ \epsilon \sum_{c\in \mathbb{D}}\sum_{l=1:m_c}\sum_{i,j\in V} t^{cl}_{ij}
 \label{eq14} \\
&\textbf{s.t.}\hspace{.5cm} \text{Constraints } (\ref{eq2})-(\ref{eq107})\nonumber
\end{align}
\begin{align}
\textit{\textbf{vars. }} \hspace{.1cm} Q,T_c,I_c,N_c\in[0,1], t_{ij}^{cl}\geq 0, \text{ and } \omega^s_{cl},\nu_c\in\{0,1\}\nonumber
\end{align}
In each time slot, FOC runs the above MILP model to maximize both QoE metrics and QoE-fairness by determining the optimal quality adaptation and data rates to transmit the requested layers. In addition to maximizing the defined QoE metrics and QoE-fairness, the objective function Eq. (\ref{eq14}) implicitly preventing the waste of network resources by decreasing \textcolor{black}{the total generated traffic (the last term in the objective function)}. We can also set desirable priorities for these metrics using constant weights $\alpha,\text{ }\beta_{1c}, \beta_{2c},\text{ and }\beta_{3c}$. For instance, $\beta_{1c}\text{ and }\beta_{2c}$ can assist us to determine how important client $c$ is to receive segments without any quality oscillation. The weights can be adjusted by the application policies in each time slot. \par
\textbf{Theorem 1:} The proposed MILP formulation (\ref{eq14}) is an NP-complete problem.

\begin{figure*}[t]
\centering
\includegraphics[width=\linewidth]{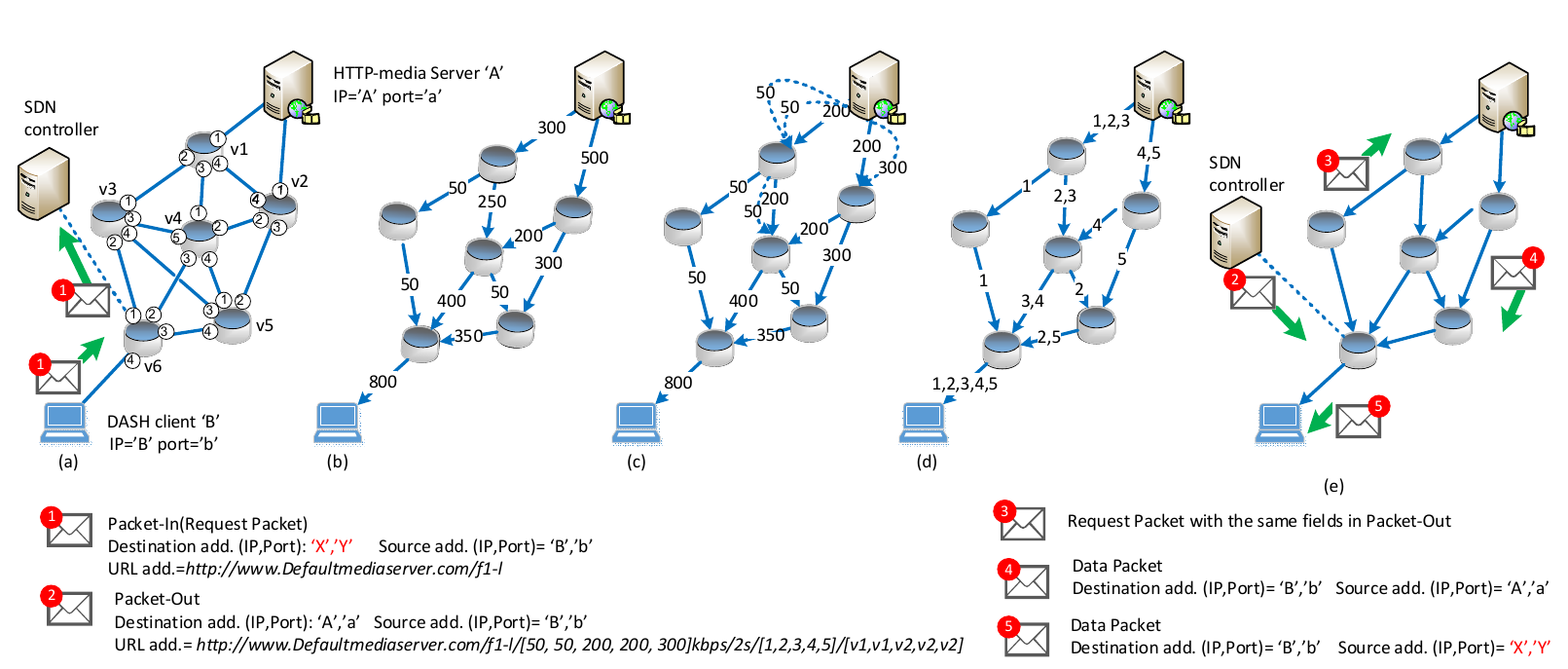}
\caption{\textbf{An example of data transmission between DASH client 'B' and HTTP-media server 'A': (a) the selected HTTP-media server and the assumed topology, (b) the determined data paths and rates by FOC, (c) obtained required tags, (d) configuring OF switches and the HTTP-media server, (e) an overview of sending control and data packets }}
\label{fig5}
\end{figure*}

\begin{proof}
Let us consider the simplest form of the problem as follows: to maximize the sum of video qualities, it is assumed that all HTTP-media servers are connected to the client-side OF switches through a link with limited bandwidth and that each HTTP-media server can serve all requested layers as well. In the simplest form, without loss of generality, we define different sets of video qualities for each segment on HTTP-media servers, where the $i^{th}$ set of video quality contains layers from the base layer to the $i^{th}$ layer. In other words, if layer $i$ is selected to be sent, then the hosted HTTP-media server will send set $i$ or all layers from the base to the $i^{th}$ enhancement layer. Now, by defining the weight and value for each stored set of video qualities in the HTTP-media servers as the required bandwidth and the perceived video quality, then the problem can be reduced in a polynomial time to the classic knapsack problem, in which the maximum knapsack capacity equals the available bandwidth. Moreover, by considering the fairness, the problem can be reduced to the quadratic knapsack problem, in which opting any two equal layers, in terms of video quality, results in more benefits.
\end{proof}

Since the proposed MILP model (\ref{eq14}) is an NP-complete problem and suffers from high time complexity \cite{vavasis1991nonlinear}, we propose a linear relaxation of the MILP model that can be centrally run by FOC. 
To do this, a virtual HTTP-media server is considered instead of set $\mathbb{S}$ by the assumption that all members of $\mathbb{S}$ are organized in a full mesh topology with no bandwidth restriction. Based on this assumption, the virtual HTTP-media server can serve all layers requested by DASH clients. Let $\chi_c\in \mathbb{R}_{\geq0}$ show the fragmentation of the maximum supported quality ($m_c$) that can be served by a virtual HTTP-media server, where $\forall c \in\mathbb{D}$ and  $\chi_c\leq m_c$. In fact,  $\chi_c$ is the amount of requested data that the virtual HTTP-media can transmit to client $c$. Now, by performing some minor modifications on (\ref{eq14}), the LP relaxation model can be represented as follows:
\begin{align}
&\textit{\textbf{minimize}} \hspace{.2cm} \alpha Q+\frac{1}{len(\mathbb{D})}\sum_{c\in\mathbb{D}}(\beta_{1c}I_c+\beta_{2c}N_c-\beta_{3c}T_c)\nonumber\\& \hspace{4.5cm}+\epsilon \sum_{c\in \mathbb{D}}\sum_{l=1:m_c}\sum_{i,j\in V} t^{cl}_{ij}
 \label{eq16} \\
 &\textit{\textbf{s.t.}} \nonumber \\
 & m_c-\chi_c\leq Q \text { }m_c , \forall c \in \mathbb{D} \hspace{4.4cm} \text{ (I)}\nonumber
   \end{align}
 \begin{align}
  &\frac{1}{\varphi_c}(\bar{\lambda}_c+\chi_c)=T_c  T_{max}  ,\forall c \in \mathbb{D}\hspace{2.8cm}\text{ (II)}\nonumber\\
 &\frac{1}{\varphi_c}(\bar{\mu}_c+\mid\chi_c-\bar{\ell}_c\mid)\leq I_c  I_{max} ,\forall c \in \mathbb{D}\hspace{1.95cm} \nonumber \text{ (III)} \\
 &\frac{1}{\varphi_c}(\bar{\nu}_c+\nu_c)\leq N_c  N_{max} ,\forall c\in \mathbb{D} \hspace{2.7cm}\nonumber\text{ (IV)}\\
 & \mid\chi_c -\bar{\ell}_c\mid\leq\nu_c  m_c ,  \forall c \in \mathbb{D}\nonumber \hspace{3.8cm} \text{ (V)}\\
 &\sum_{c\in \mathbb{D}} e_{ij}  t^{c}_{ij}\leq b_{ij} , \forall \text{ } i,j \in V\nonumber \hspace{3.6cm} \text{ (VI)}\\
 &\hspace{.1cm}\theta_c(\sum_{j\in V} e_{ij}  t^{c}_{ij}-\sum_{j\in V}e_{ij}  t^{c}_{ji})=\nonumber\\
&\left\{\begin{array}{ll}\bar{\delta_{c}} \chi_c \hspace{2cm},\forall i \in \mathbb{S}, c\in\mathbb{D} \\0\hspace{1cm},\forall i \in V-\{\mathbb{S},\mathbb{D}\},\text{ }c\in\mathbb{D}\hspace{2cm} \text { (VII)}\\-\bar{\delta}_{c} \chi_c\hspace{1.6cm},\forall \text{ } i\in\mathbb{C},\text{ }c\in\mathcal{N}_i\\\end{array}\right.\nonumber 
\end{align}
\begin{align}
 &\textit{\textbf{vars. }}  \hspace{1cm} \chi_c,t^c_{ij}\geq0,\textit{ }Q,T_c,I_c,N_c,\nu_c\in[0,1]\nonumber
 \end{align}
In constraints (VI-VII), $t^{c}_{ij}$ is defined as the data rate at which $i \in V$ sends video traffic requested by  $c \in \mathbb{D}$ to $j \in V$ in $\theta_c$ units of time. Also, $\bar{\delta}_c$ is the average layer size of the segment requested by client $c$.
By taking the floor of the fractional values of $\chi_c$, an approximate solution to the model (\ref{eq14}) can be obtained. After determining the sub-optimal data rates and paths, the flow setup component is called.\par

\textbf{Flow Setup Component (FSC)}: After determining the sub-optimal solution by FOC, a method should be developed and utilized to enable the network to support different data transmission rates. To achieve this, OF switches and HTTP-media servers should be configured accordingly. 



As stated in \cite{openvswitch}, Open vSwitch (OVS) is able to provide desirable data rate calibration on each flow using its queuing feature. However, configuring different queues on all OF switches may be a simple, yet naive idea because of two reasons. First, there are numerous video files with different data rates that must be delivered in each time slot. Second, the probability of misconfiguration grows as the number of switches increases.

Therefore, we propose an agile and reliable method based on tagging the \textit{type of service} (TOS) field of the data packets originated by the HTTP-media servers to the destination DASH clients. Generally, the tagging process can be performed by HTTP-media servers or server-side OF switches. However, to achieve a minimum overhead in OF switches, we delegate the tagging process to the HTTP-media servers. \par
The FSC operation for each requested layer can be divided into two main steps: (1) configuring the HTTP-media server to tag the outgoing packets, and (2) configuring OF switches to provide the determined data path and rates.
\par
\textbf{Step 1:} In this step, FSC should determine the minimum required tags for the generated data packets and then configure the selected HTTP-media server to apply the tags. For ease of explanation, let us present an example illustrated in Fig. \ref{fig5}. Suppose all DASH clients are configured to send their request to a default HTTP-media server with IP='X' and port='Y'. Thus, as illustrated in Fig. \ref{fig5} (a), the client-side OF switch '$v6$' receives a request from DASH client '$B$' with destination IP 'X' and port 'Y'. Then, it forwards this packet to the SDN controller as a PACKET-IN message. Upon receiving the PACKET-IN message, the SDN controller delivers it to its RAC component for analyzing the packet. After processing the request received by RAC, FOC selects the HTTP-media server '$A$' to serve DASH client '$B$' for layer '$l$' based on the optimal data path and transmission rates, (see Fig. \ref{fig5} (b) for the given deadline $\theta_B=2$ seconds). Now, with the optimal data path and rate $t^{Bl}_{ij}$, the minimum number of tags must be calculated by FSC. In fact, a list of \textit{base} data rates $\mathbb{B}_{Bl}=\{b_1, b_2, ...,b_n\}$ must be produced, in which $\forall i,j\in V$, $t^{Bl}_{ij}$ is decomposing into $\mathbb{B}_{Bl}$. To determine the $\mathbb{B}_{Bl}$ values, we propose a simple recursive function named \textit{base\_data\_rates} (see Alg. \ref{alg1}).\par

\begin{algorithm}[t]
    \caption{base\_data\_rates(n, m) }
    \label{alg1}
    \SetAlgoLined
    //$ \textit{m=server '$A$' and in the first call n= switch '$v6$' }$\;
    \If {$n==m$} {
        return\;
        }
    $L$=all\_incoming\_link($n$)\;
    \ForEach {$l$ in $L$} {
    [$n$,$p$] = peers($l$)\; 
    $\bar{L}$ = unvisited\_incoming\_link($p$)\; 
    \If {data\_rate($l$) = any\_combination(data\_rate($\bar{L}$))}
    {
    set the true combination of $\bar{L}$ as visited\;
    continue\;
    }
    \eIf{data\_rate($l$) $>$ max\_rate($\bar{L}$)}
    {
    Call update\_data\_rate\_I($l$, $\bar{L}$)\;
    }
    {
    Call update\_data\_rate\_II($l$, $\bar{L}$)\;
    }
    $U$ = upstream\_OFswitch($n$)\;
    \ForEach{$u$ in $U$}
    {
    Call base\_data\_rates($u,m$)\;
    }
    }
\end{algorithm}

This algorithm traverses the data path from the client-side OF switch '$v6$' to HTTP-media server '$A$' by calling function $base\_data\_rates$ ('$v6$','$ A$'). This function stops calling itself when it meets the root of data path (HTTP-media server '$A$') (lines 1-3). In line 4, list $L$ is set to all incoming links of $n$ (i.e. $v3\rightarrow v6$, $v4\rightarrow v6$, and $v5\rightarrow v6$ for $n=v6$).
For each link $l$ in $L$, we first define $\bar{L}$ as the list of  incoming links to $p$ that have not been visited yet, where $p$ is the upstream peer for $n$ on link $l$. For instance, for $n=v6$, $l=v4\rightarrow v6$, $p=v4$, and $\bar{L}=\{v1\rightarrow v3\}$, if the algorithm finds a combination of data rates of $\bar{L}$ equal to the data rate of $l$, then all links in that combination are set as visited links and the for loop goes for the next link (lines 9-12). Consider Fig. \ref{fig5}(b) as an example, in which the condition of the If statement (line 9) is TRUE for link $l=v3\rightarrow v6$, since  $data\_rate(l)=data\_rate(v1\rightarrow v3)$. As it can be seen in Fig. \ref{fig5}(b), for $n=v6$, the data rate from $v3$ to $v6$ is equal to the data rate from $v1$ to $v3$, hence $v1\rightarrow v3$ is set as the visited link. 

However, for $n=v6$ and $l=v4\rightarrow v6$, there is not any combination of $\bar{L}=\{v1\rightarrow v4, v2\rightarrow v4\}$ that provides the same data rate as $v4\rightarrow v6$. In this case, we compare $data\_rate(l)$ with the maximum data rate of $\bar{L}$ (here $v1\rightarrow v4$ with 250 kbps). If $data\_rate(l)>max\_rate(\bar{L})$, then the function $update\_data\_rate\_I(l.\bar{L})$ (Alg. \ref{alg2}) is called. In this algorithm, a combination of unvisited links to $p$ (e.g., $v4$ for $n=v6$) is determined firstly, which provides the minimum data rate that is greater than $data\_rate(l)$ (lines 4-9). Then, virtual link $vl$, which is added to $\bar{L}$, and the extra data rate are allocated to it (lines 10-13). Refer to link $v4\rightarrow v6$ in Fig. \ref{fig5} (c) which results in the addition of a virtual link $v1\rightarrow v4$ with data rate 50. 

On the other hand, in the case of $data\_rate(l)<max\_rate(\bar{L})$, the next function, $update\_data\_rate\_II(l.\bar{L})$, executes (Alg. \ref{alg3}). Let $l2$ have the maximum data rate of $\bar{L}$ (for $n=v4$ and $l=v2\rightarrow v4$, we have $l2=HTTP-media server$ '$A$'$\rightarrow v2$). In this case, virtual link $vl$ with the same data rate $data\_rate(l)$ is added to  $\bar{L}$. We also set $vl$ as the visited link (line 4). Since the data rate of $vl$ must equal $data\_rate(l)$, we then update the data rate of $l2$  to $data\_rate(l2)-data\_rate(l)$ (line 6). See link $v2\rightarrow v4$ in Fig. \ref{fig5} (c) which results in the addition of a virtual link from HTTP-media server '$A$' to $v2$ with a data rate of 300). In line 20 of the $base\_data\_rates (n,m)$ algorithm, the function calls itself for each $n$'s upstream switches. Finally, the list of base data rates $\mathbb{B}_{Bl}$ is equal to the all outgoing data rates, including the data rates of virtual links. In this example, the obtained base data rates are $\{50kbps, 50kbps, 200kbps, 200kbps, 300kbps\}$ (refer to the outgoing links from HTTP-media server '$A$' in Fig. \ref{fig5} (c)).

\begin{algorithm}[t]
    \caption{update\_data\_rate\_I($l$, $\bar{L}$) }
    \label{alg2}
    \SetAlgoLined
    $l2$ = find\_link(min\_rate($\bar{L}$))\;
    set $l2$ as visited\;
    $r$=data\_rate($l2$)\;
    \While{$r <$ data\_rate($l$)}
    {
        delete\_link($\bar{L}$,$l2$)\;
        $l2$=find\_link(min\_rate($\bar{L}$))\;
        set $l2$ as visited\;
        $r$ = $r+$ data\_rate($l2$)\;
    }
    [$p$,$q$]=peers($l2$)\;
    $vl$= add\_virtual\_link($p$,$q$)\;
    set\_data\_rate($vl$, $r-$data\_rate($l$))\;
    update\_data\_rate($l2$, data\_rate($l2$) $-$ ($r-$ data\_rate($l$)))\;
     \end{algorithm}
     \begin{algorithm}[t]
    \caption{update\_data\_rate\_II($l$, $\bar{L}$) }
    \label{alg3}
    \SetAlgoLined
        $l2$ = find\_link(max\_rate($\bar{L}$))\;
        [$p$,$q$] = peers ($l2$)\;
        $vl$ = add\_virtual\_link($p$,$q$)\;
        set $vl$ as visited\;
        set\_data\_rate($vl$,data\_rate($l$))\;
        update\_data\_rate($l2$,  data\_rate($l2$)$-$data\_rate($l$))\;
\end{algorithm}

After the base data rates are calculated, FSC configures the HTTP-media server to tag the outgoing packets accordingly. To do this, FSC concatenates the following parameters to the URL address in the request packet: (1) base data rates, (2) deadline value $\theta_c$, (3) list of required tags, and (4) list of server-side OF switches connected to the selected HTTP-media server. Thereafter, FSC asks the SDN controller to send the outgoing packets back to the client-side OF switch as a PACKET-OUT message. Note that the destination address of the PACKET-OUT message is set to the selected HTTP-media server. The client-side OF switch now forwards the packet to the specified HTTP-media server (Fig. \ref{fig5} (e)). \par 
As depicted in Fig. \ref{fig5} (d)-(e), for the obtained base data rates (i.e., $\{50kbps, 50kbps, 200kbps, 200kbps, 300kbps\}$), the FSC sets $tag=\{1,2,3,4,5\}$ and \textit{server-side OF switches}=$\{v1,v1,v1,v2,v2\}$. This configuration signifies that the HTTP-media server '$A$' must send the amount of $\theta_B\times50kbps$ data to server-side OF switch $v1$ with tag 1, the amount of $\theta_B\times50kbps$ data to $v1$ with tag 2, and so forth.\par

\textbf{Step 2:}
In this step, FSC installs the required rules in OF switches to forward packets according to the determined data path. Each rule filters incoming packets according to the ToS tag, source and destination IP and port addresses, and then specifies the output port (see Fig. \ref{fig5}(d)). For more transparency, the client-side OF switch replaces the source IP and port addresses of the incoming data packet from the HTTP-media server with the default IP and port addresses. (Fig. \ref{fig5}(e)).

\begin{figure}
\centering
\includegraphics[width=\linewidth]{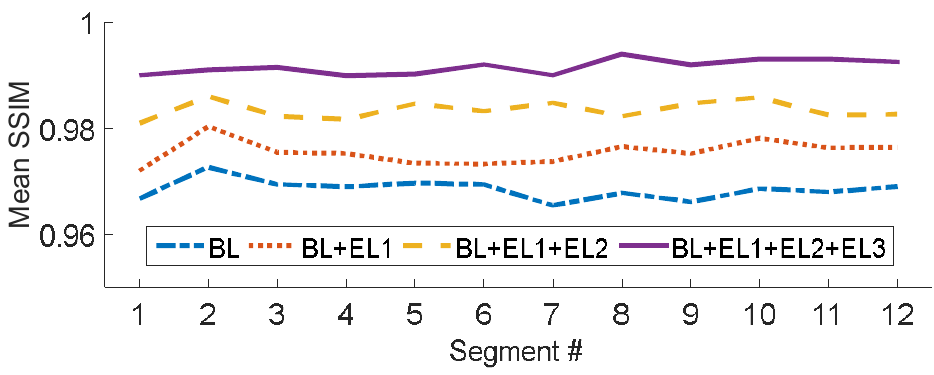}
\caption{\textbf{Measured video qualities with SSIM \cite{wang2004video}}}
\label{P_100}
\end{figure}

\section{Performance Evaluation}
In this part, we empirically evaluate and present the performance results of the proposed models. In the following, we first investigate the behavior of the proposed MILP model and various QoE parameters. Then, we continue our investigation by comparing the performance of proposed QoE-fairness method with that of other algorithms. Finally, we evaluate the performance of the proposed MILP and LP-relaxed models in detail.\par

For evaluation purposes, we use the 56-second \textit{Factory.yuv} obtained from \textit{https://media.xiph.org/video/derf/}. The video is processed using the Joint Scalable Video Model (JSVM) \cite{reichel2007joint} as follows. First, the video is encoded by \textit{JSVM H264AVCEncoderLibTestStatic} to four scalable layers, including temporal and spatial scalability. The encoder produces one .264 file. We use the Python code presented in \cite{kreuzberger2015scalable}  to extract the corresponding file to each layer and  slice this into five-second long segments. The characteristics of the produced file are as follows: base layer (BL)=650 kbps, BL+enhancement layer1 (EL1)=1100 kbps, BL+EL1+EL2=1650 kbps, and BL+EL1+EL2+EL3=2300 kbps. Using SSIM\cite{wang2004video}, we measure the video qualities of layers and illustrate in Fig. \ref{P_100}. Then, the related MPD file is compiled based on the extracted layers information. Finally, the created files are uploaded into simulated web servers in Mininet \cite{lantz2010network}. 

\begin{figure}[t]
\centering
  \centering
\includegraphics[width=.8\linewidth]{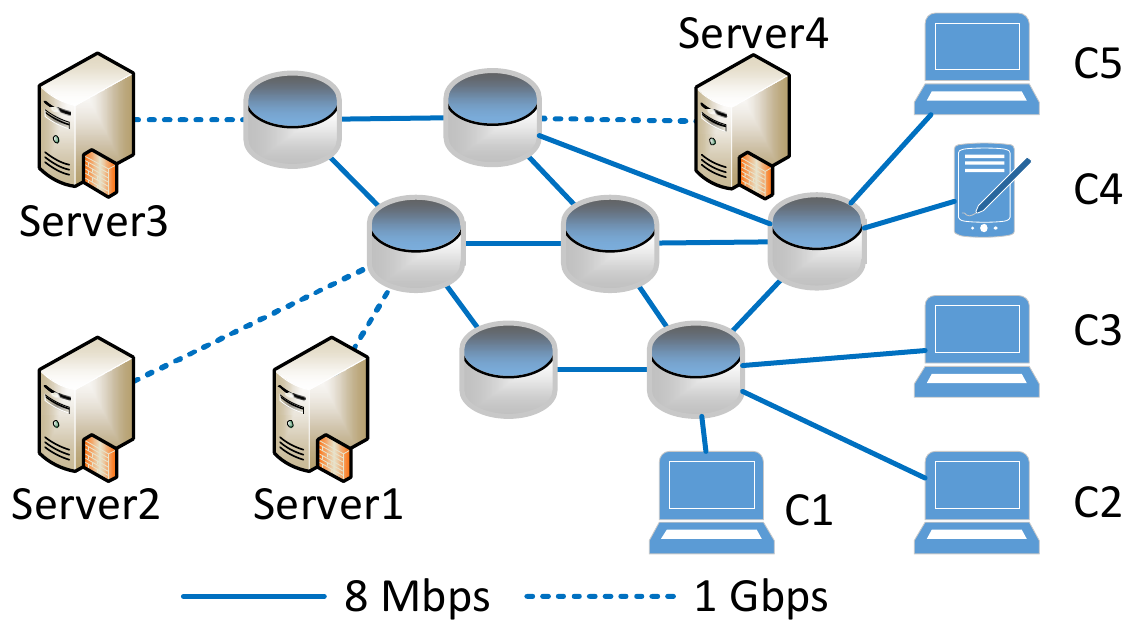}
\caption{\textbf{The default topology}}
\label{fig66}
\end{figure}

The default topology used in the implementation is illustrated in Fig. \ref{fig66}. We used computers running Ubuntu 16.04 64bit and equipped with an Intel Core 2 Duo CPU and 4 GB of RAM. There are seven OF switches and five clients connected to the system via two client-side OF switches. As seen in Fig. \ref{fig66}, the bandwidth between the switches is set at 8 Mbps. In fact, to simulate a bandwidth-limited network and to investigate the treatment of the framework, we limit the core network bandwidth capacity (i.e. switch to switch). Furthermore, the test bed consists of five heterogeneous clients $C_1-C_5$ with different supported layers: $m_1=4,\text{ }m_2=4,\text{ }m_3=3,\text{ }m_4=2,\text{ and }m_5=4$, respectively. In our setup, clients join the network in predefined random time-slots, as follows: $C_1=1,\text{ }C_2=4,\text{ }C_3=6,\text{ }C_4=6,\text{ and }C_5=6$. Thereafter, they request receiving the 56-second video in 12 segment, each segment containing 5 seconds of the video. The other initial parameter values are set as follows:  $\theta_c=1s,\tau=2s,\alpha=1,\epsilon=0.1,\beta_{1c}=0.2,\beta_{2c}=0.2,\text{ and }\beta_{3c}=1 \text{ }(for c=1:5)$.
We use Mininet to generate the network topology and perform the evaluations. We also employ Floodlight \cite{floodlight} as the SDN controller in our experiments. To emulate the clients' video player, we extend \textit{Scootplayer} \cite{scoot} to support the H.264 SVC codec. \textit{Scootplayer} is an experimental MPEG-DASH request engine which also provides accurate logs. Next, we evaluate the behavior of the system by changing the values of different parameters.

\subsection{QoE Parameters}
According to the buffering deadline '$\theta_c$' and the requested segment duration, we can eliminate the stalling phenomenon in clients by setting $\theta_c\leq$ $(segment\_duration) - (time\_slot\_duration$ $\tau)$. The transmission delay can also be considered when determining the $\theta_c$ value. Thus, by choosing $\theta_{c} \leq 3s$, model can guarantee that stalling phenomenon never happened. The next QoE parameter, start-up delay, depends on $\tau$ and $\theta_c$ too. 
In fact, in the worst case, a client may join the network exactly at the beginning of when the optimization algorithm is run. In this case, the client has to wait for $\tau + \theta_c$ seconds to receive the video. On the other hand, in the best case, the client's request is submitted exactly before the optimization algorithm is run and the video is delivered after $\theta_c$ seconds. Notably, the average waiting time for clients may be expressed as $\theta_c + \frac{\tau}{2}$. The empirical results proved these statements.

\begin{figure}[t]
\centering
\includegraphics[width=\linewidth]{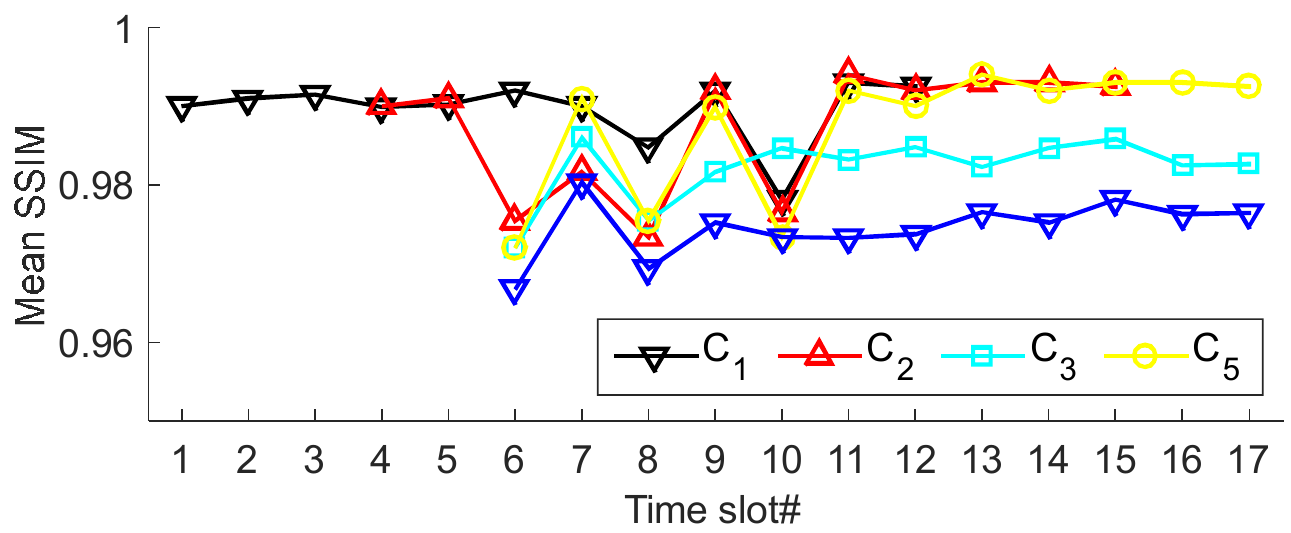}
\caption{\textbf{Measuring the mean SSIM \cite{wang2004video} for clients}}
\label{p1_ssim}
\end{figure}

\begin{figure}[b]
\centering
\includegraphics[width=\linewidth]{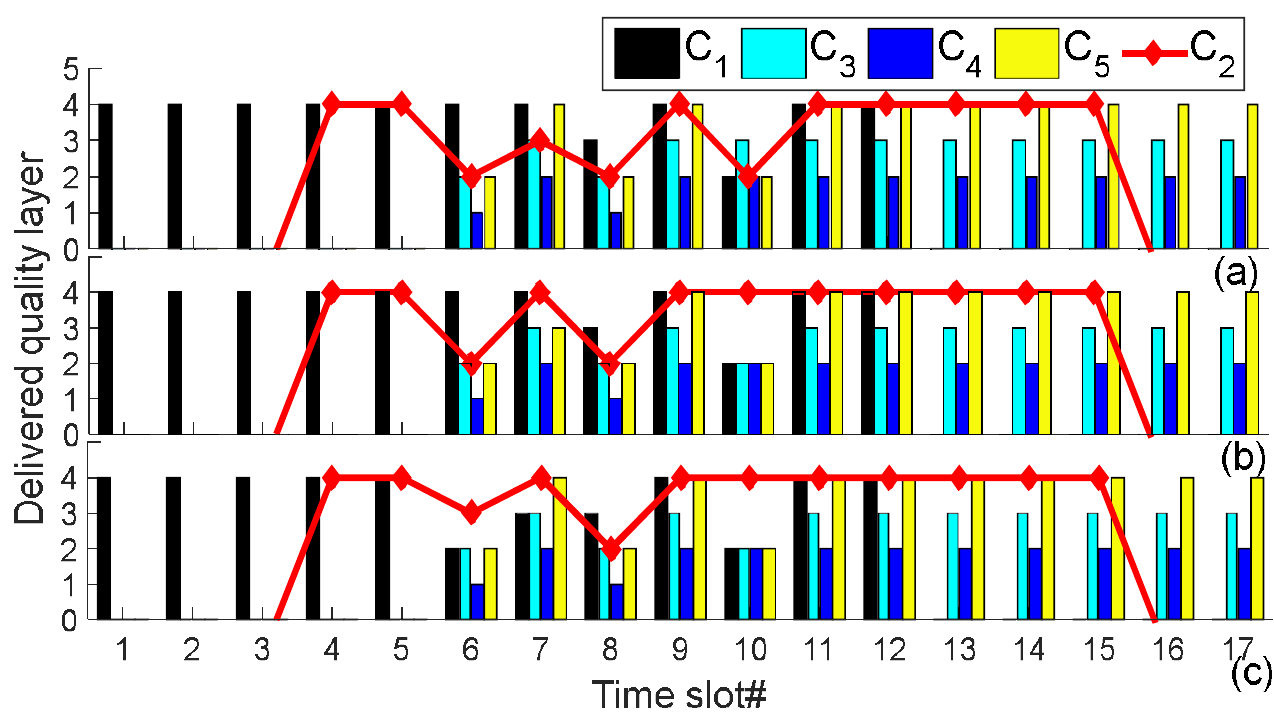}
\caption{\textbf{The impact of changing $\beta_{3c}$ value: (a)$\beta_{32}=1$, (b) $\beta_{32}=2$ and (c) $\beta_{32}=3$}}
\label{code234}
\end{figure}

In the first experiment, we focus on the video quality received by clients. To measure this parameter, we employ the Structural Similarity Metric (SSIM) \cite{wang2004video}. In fact, SSIM can be utilized to assess the mean delivered video quality to clients in each time slot (Fig. \ref{p1_ssim}). As mentioned in the proposed model, we set the video quality of a segment equal to the number of its received layers. Now, by comparing Fig. \ref{p1_ssim} and Fig. \ref{code234}(a), it can be concluded that both methods follow the same approach. Thus, this proves that considering the number of received layers for a segment can be mapped as an index for measuring video quality. In our next experiment, we measure the changing weight of client 2 to show the capability of the system to apply different policies.

We therefore evaluate the impact of changing the $\beta_{3c}$ value for $c=2$. This value is related to parameter $T_c$ in model (\ref{eq14}), which shows the average video quality of each client $c$ from the beginning to the present time. With the increase of $\beta_{32}$, it is expected that client 2 receives a higher video quality in comparison to others. In Fig. \ref{code234}(a), the behavior of the system with the default value of $\beta_{32}$ is depicted. As shown in Fig. \ref{code234}(b), by doubling the value of this parameter, the delivered video quality for client 2 increases in time slots 7 and 10. Moreover, by considering the bandwidth limitation, it can be seen that, due to the higher video quality of $C_2$, the video quality of $C_3$ and $C_5$ lowers. When bringing the $\beta_{32}$ value further up to 3. Fig. \ref{code234}(c) shows that the video quality of $C_2$ improves in time slot 6. Therefore, by raising the $\beta_{32}$ value to higher values, we can guarantee the maximum quality for client 2 during all time slots.\par

\begin{figure}[t]
\centering
\includegraphics[width=\linewidth]{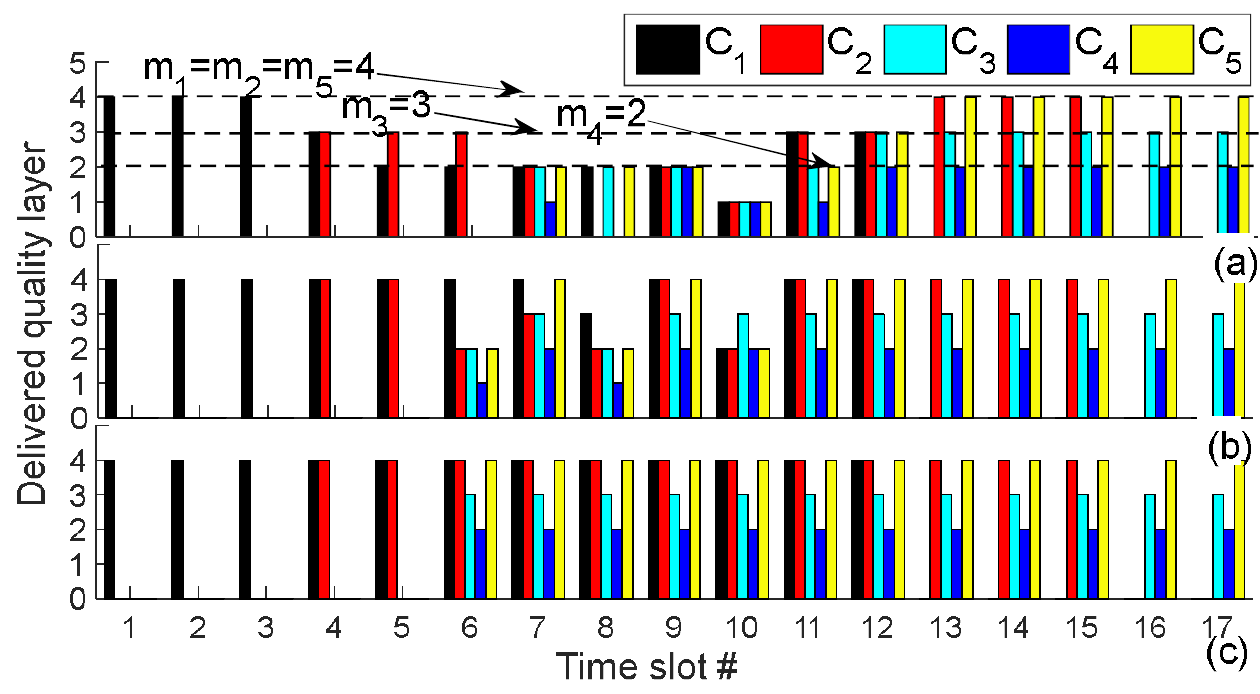}
\caption{\textbf{The impact of changing $\theta_c$ value: (a) $\theta_c=0.6$, (b) $\theta_c=1$, and (c) $\theta_c=1.7$}}
\label{code11}
\end{figure}
\begin{figure}[b]
\centering
 \includegraphics[width=\linewidth]{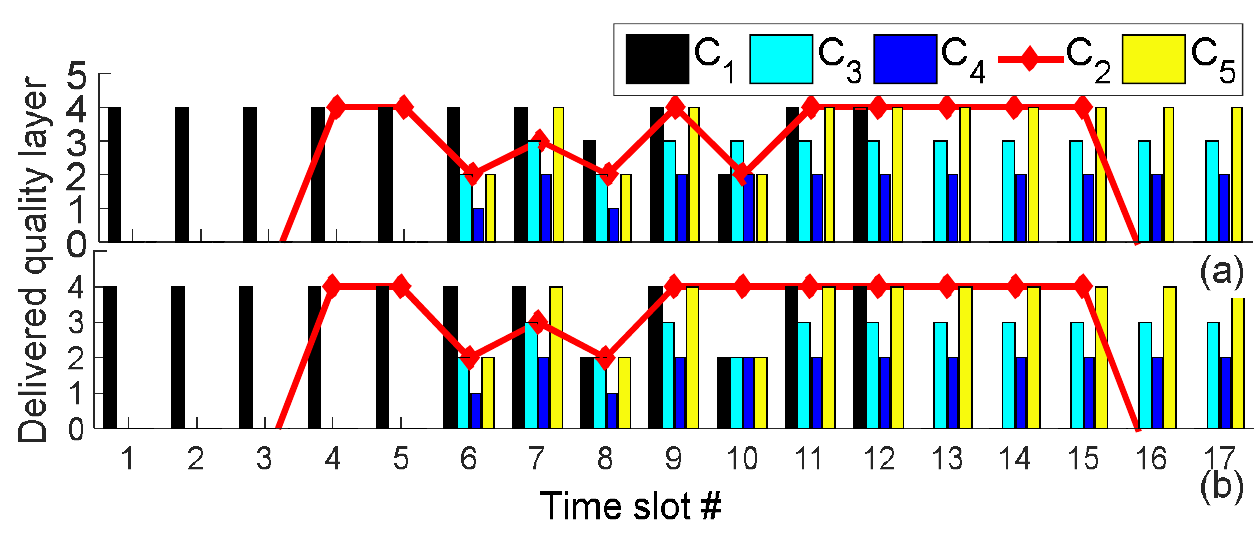}
\caption{\textbf{The impact of changing $\beta_{12}$ and $\beta_{22}$ values: (a)$\beta_{12}=\beta_{22}=0.2$ and (b) $\beta_{12}=\beta_{22}=1$ }}
\label{code56}
\end{figure}

In the next experiment, we investigate the impact of changing $\theta_c$ in the delivered quality layers. As a reminder, $\theta_c$ is defined as the video delivery deadline for client $c$. $\theta_c$ is assumed to be identical for all clients. Obviously, by increasing the value of $\theta_c$, resource utilization can decrease and a higher video quality may be delivered to clients. This is because the proposed model has more time to deliver the requested video and consequently it is able to send higher video quality at a lower transmission data rate. The treatments of the model for different values of $\theta_c$ are illustrated in Fig. \ref{code11}. The normal status of the network ($\theta_c=1$) is shown in Fig. \ref{code11} (b). As seen, by increasing the $\theta_c$ value from 1 to 1.7, the maximum video quality (i.e. $m_c$) can be delivered to clients (see Fig. \ref{code11} (c)). Although raising $\theta_c$ promotes the overall quality of the delivered video, the possibility of stalling occurring for clients grows. In fact, the stalling phenomenon specifies an upper bound for $\theta_c$. Basically, the proposed framework can centrally offer an appropriate $\theta_c$ by considering the total duration of delivering video to client $c$. We will elaborate on this in a future study. Notably, by choosing a smaller value for $\theta_c$ (Fig. \ref{code11} (a)), we can force the model to immediately fill the buffer of the clients; in this case, the model has to send the highest possible quality to the clients at high data rate. Therefore, as for limited bandwidth, it is obvious that some clients will be deprived of obtaining more quality (see Fig. \ref{code11} (a) for $\theta_c=0.6$).\par
In the next experiment, we evaluate the impact of changing the values of the other two weights in our model: $\beta_{1c}$ and $\beta_{2c}$. By increasing the values of these weighting factors for $C_2$, it is expected that the number and intensity of quality changes shall decrease. Fig. \ref{code56} (a) presents the system behavior with the default settings. In the first step, we increase the value of $\beta_{12}$ and $\beta_{22}$ to $1$. As seen in Fig. \ref{code56}(b), the controller tries to reduce number of the quality changes, which increases the video quality in time slot 10. Also, to raise the video quality of $C_2$, the controller has to reduce the video quality of other clients. For example, in the 10th time slot, the MILP model lowers the quality of $C_3$ to boost the video quality of $C_2$ (see Fig. \ref{code56} (b)).\par
As presented in Fig. \ref{code234} (a) and \ref{code56} (a) (default cases), the requested quality was not completely delivered in time slots 6, 7, 8 and 10. To clarify the reason for this treatment, we measure the total volume of requested layers by the clients and the amount of sent data by MILP in each time slot. In fact, due to the bandwidth limitation, the MILP model can not deliver the total amount of requested layers in some time slots 6, 7, 8, and 10. Although the number of delivered quality layers lowered in time slots 6, 8 and 10, the amount of sent data in these slots is reasonable when considering the bandwidth limitation. This phenomenon occurs in other scenarios (see Fig. \ref{code234} (b,c) and \ref{code56} (b)).
\subsection{QoE-Fairness}
\begin{figure}[t]
\centering
\includegraphics[width=\linewidth]{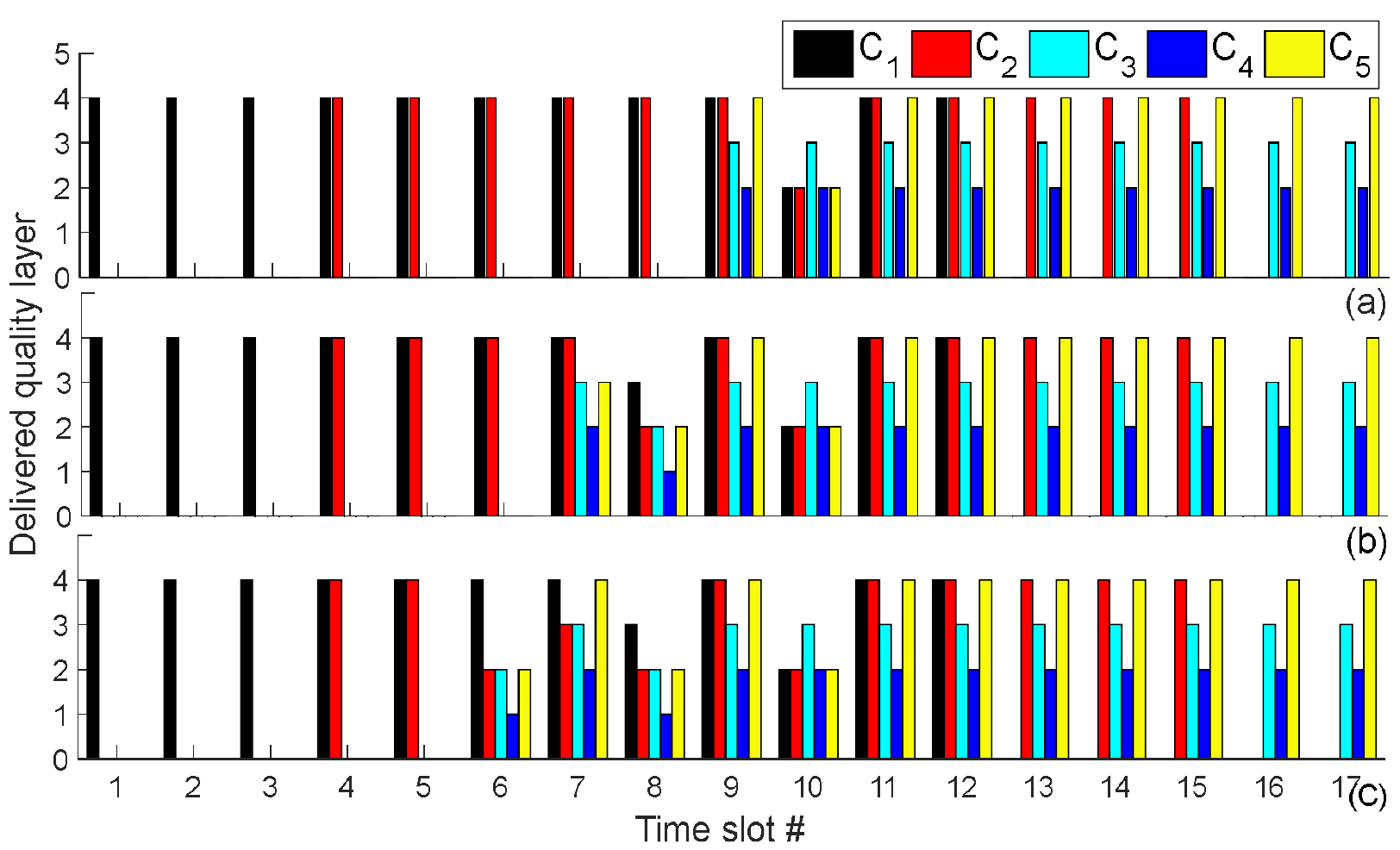}
\caption{\textbf{The impact of changing $\alpha_{1}$ on the delivered quality layers: (a)$\alpha_{1}=0.1$, (b)$\alpha_{1}=0.5$, and (c)$\alpha_{1}=1$ }}
\label{code78}
\end{figure}

Parameter $\alpha$ is responsible for providing video quality fairness in the network (see the proposed model (\ref{eq14})). Fig. \ref{code78}(c) presents the status of the system in default mode ($\alpha=1$). As depicted in Fig. \ref{code78}, by decreasing the value of $\alpha$ from $1$ to $0.5$ and also lowering it to $0.1$, the MILP model does not try to serve all requested clients.

\begin{figure}[b]
\centering
\includegraphics[width=\linewidth]{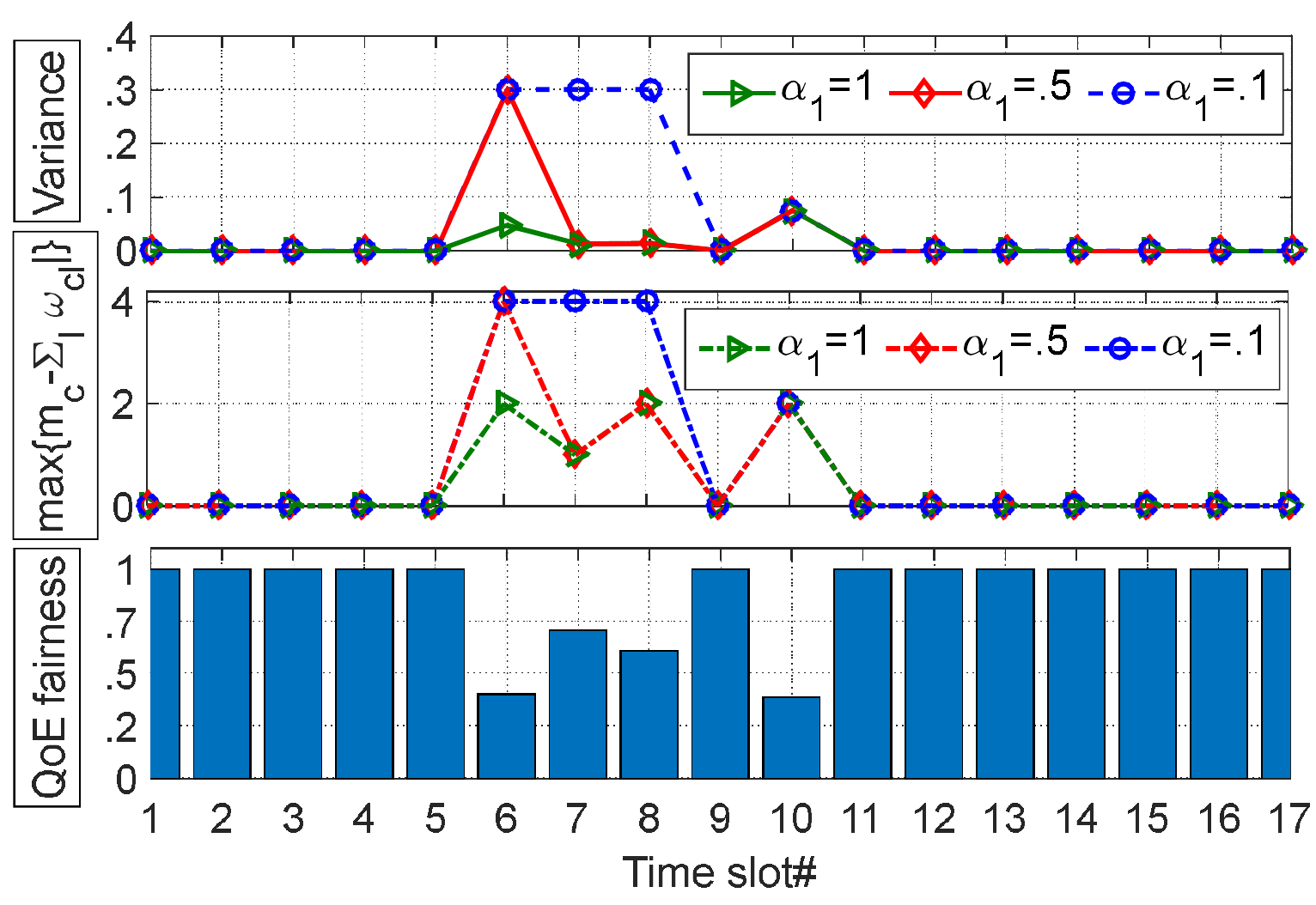}
\caption{\textbf{The impact of changing $\alpha_{1}$ value on the fairness}}
\label{code910}
\end{figure}

For example, in the cases of $\alpha=0.5$ (Fig. \ref{code78} (b)) and $\alpha=0.1$ (Fig. \ref{code78} (a)), the Flow Optimizer component (FOC) decides to respond to $C_3-C_5$ after the 6th and 8th time slots, respectively. As a result, lowering $\alpha$ can reduce the overall QoE-fairness. Moreover, by considering different values for  $\alpha$, we measure the variance of normalized received quality layers in each time slot and also the maximum gap between the maximum supported video quality layer $m_c$ and the delivered layer that is equal to $max\{m_c-\sum_l\omega_{cl}\mid\forall c\}$ (refer to Fig. \ref{code910}). As seen, this difference grows dramatically in the 6th time slot for $\alpha=0.1$ and $\alpha=0.5$; however, it reduces after $\alpha$ increases. An identical treatment showing variance in Fig. \ref{code910}. Furthermore, we use the QoE-fairness index introduced by Hossfeld et al. \cite{hossfeld2017definition} as follows:
\begin{equation}
    F_{SSIM}=1-\frac{2\times \sigma_{SSIM}}{\sigma_{Max}-\sigma_{Min}}
\end{equation}

\begin{figure}[t]
    \centering
    \includegraphics[width=.75\linewidth]{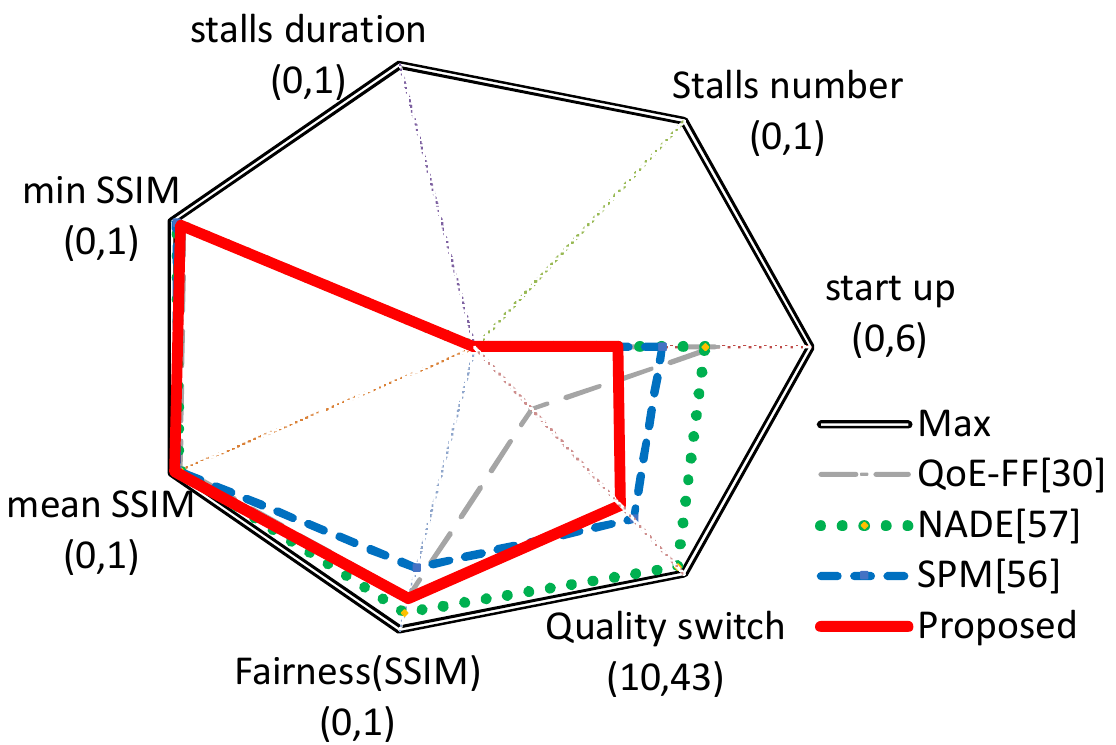}
    \caption{\textcolor{black}{\textbf{Comparing the proposed MILP model with other approaches}}}
    \label{p2_3}
\end{figure}

\noindent where $\sigma_{SSIM}$ denotes standard deviation of SSIM values observed by clients in each time slot. $\sigma_{Max} \text{ and } \sigma_{Min}$ represent the maximum and minimum SSIM values for available videos in HTTP media-servers, respectively.
As depicted in Fig. \ref{code910}, our measured QoE-fairness and that measured by \cite{hossfeld2017definition} share the same trend.\par 
By employing the proposed framework in \cite{schwarzmann2017quantitative}, we compare our introduced MILP model with three network-based methods: QoE-FF framework\cite{georgopoulos2013towards}, SPM\cite{petrangeli2015network}, and NADE\cite{cofano2016design}. As it is depicted in Fig. \ref{p2_3}, our proposed model is able to outperform other approaches in term of start-up delay. NADE proposed a video control plane which enforces video quality fairness among concurrent video flows generated by heterogeneous client devices. Also, it can be seen that NADE yields the highest fairness among all the approaches. Nevertheless, our proposed model performs almost the same as NADE. On the other hand, QoE-FF outperforms NADE, SPM, and our proposed model in term of quality switches.  The reason is NADE is not designed to foresee the occurrence of video freezes and avoid them. However, as it appears in Fig. 14, our proposed model produces acceptable results for this parameter. QoE-FF seeks to optimize the QoE by taking into account two main constraints: the devices resolution and current available bandwidth. However, it does not consider the current buffer occupancy. Thus, for both QoE-FF and NADE mechanisms, decreasing the available bandwidth can significantly increase the stalling phenomenon. As a result, in these approaches, clients may be subject to buffer starvation. Meanwhile, our proposed model guarantees that stalling will never happen. Moreover, QoE-FF framework does not support large number of clients, since it generates high overhead which can degrade the overall performance. Finally, Fig. 14 shows that all the studied frameworks produce almost the same results for both min and mean SSIM.
\begin{figure}[t]
\centering
 \includegraphics[width=.9\linewidth]{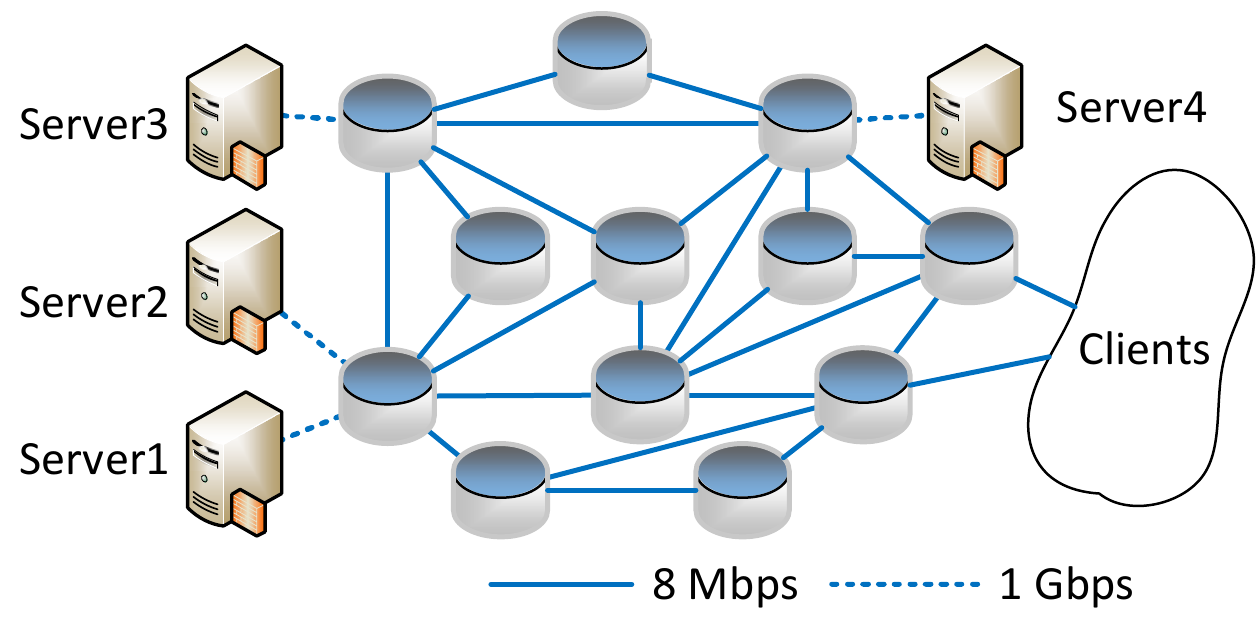}
\caption{\textbf{The extended topology}}
\label{code13151617} 
\end{figure}
\begin{table}[t]
    \centering
    \captionsetup{justification=centering}
    \caption{ \textbf{The MILP and LP-relaxed comparison \newline in default(D) and extended(E) topology}}
\begin{tabular}{c |c| c| c |c|c|c }
  \hline
   \multicolumn{3}{c|}{Number of}& \multicolumn{4}{c}{Execution time (ms)}
   \\
      Client & \multicolumn{2}{c|}{OVS command} & \multicolumn{2}{c|}{MILP} & \multicolumn{2}{c}{LP-relaxed} \\
       & D& E&D&E&D&E\\\hline
         1 & 17&17&80&130&35&55 \\
        5&87&91&311&510&89&180\\
         10&177&177&743&1700&173&293\\
        15&220&225&1043&2250&249&458\\
        20&305&321&2152&3500&296&630\\
        25&328&345&3748&5200&336&764\\
        30&368&398&4447&6800&350&837\\
        40&442&514&6276&8860&415&1029\\
\hline
\end{tabular}
\label{T4}
\end{table}

\begin{figure}[b]
\centering
 \includegraphics[width=\linewidth]{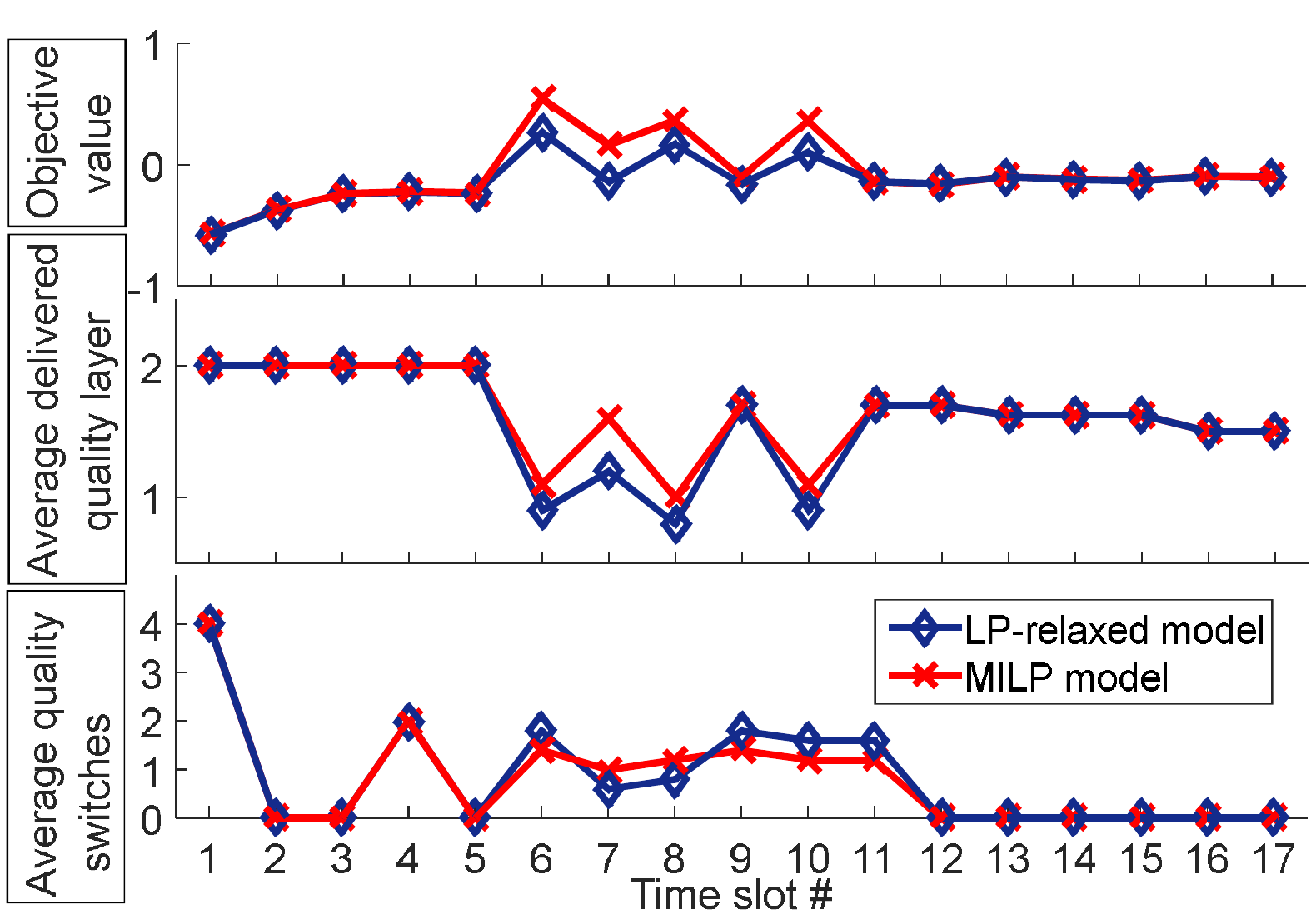}
\caption{\textbf{The performance comparison between the LP-relaxed and MILP models}}
\label{code131516} 
\end{figure}

\subsection{MILP and LP-Relaxed Comparison}
Before discussing the comparison of the proposed models, let us provide a brief overview of the number of generated OF commands for configuring the network switches. \textcolor{black}{We conduct experiments with default (D) topology (Fig. \ref{fig66}) then enrich our experiments with extended (E) topology (Fig. \ref{code13151617}) (by increasing the number of switches to 12) and  measure the number of OF commands for all the active switches in each time slot.} After a traffic flow finishes, each flow table entry will be removed from the switch. As seen in Table \ref{T4}, by increasing the number of clients, the number of OF commands moderately increases, because of dividing traffic and utilizing more data paths to the destination. \par
\begin{figure}[t]
   \centering
\includegraphics[width=\linewidth]{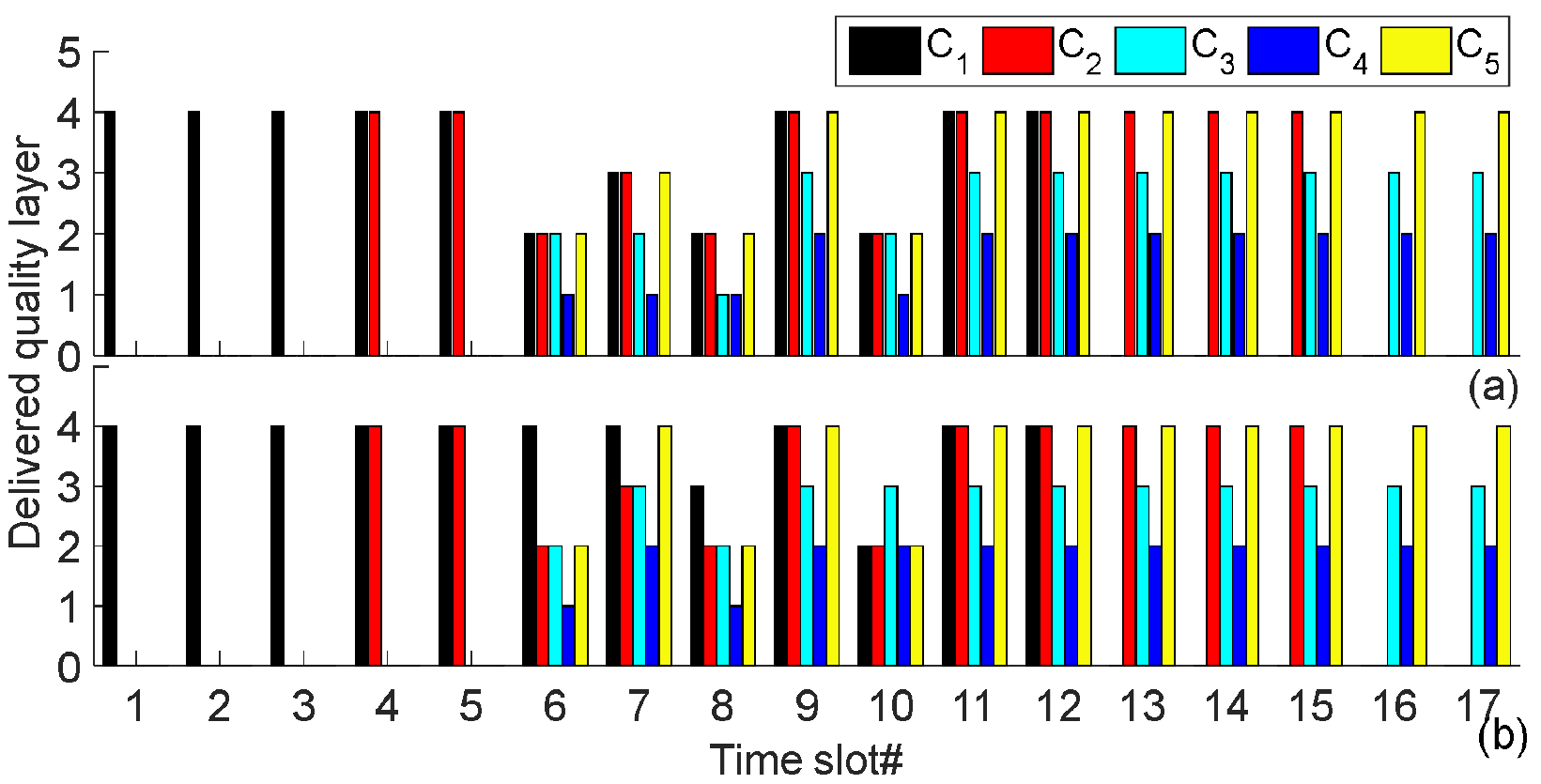}
\caption{\textbf{Delivered quality layer in each time slot: (a) LP-relaxed model and (b) MILP model}}
\label{figg1555}
\end{figure}

\begin{figure}[b]
\centering
\includegraphics[width=\linewidth]{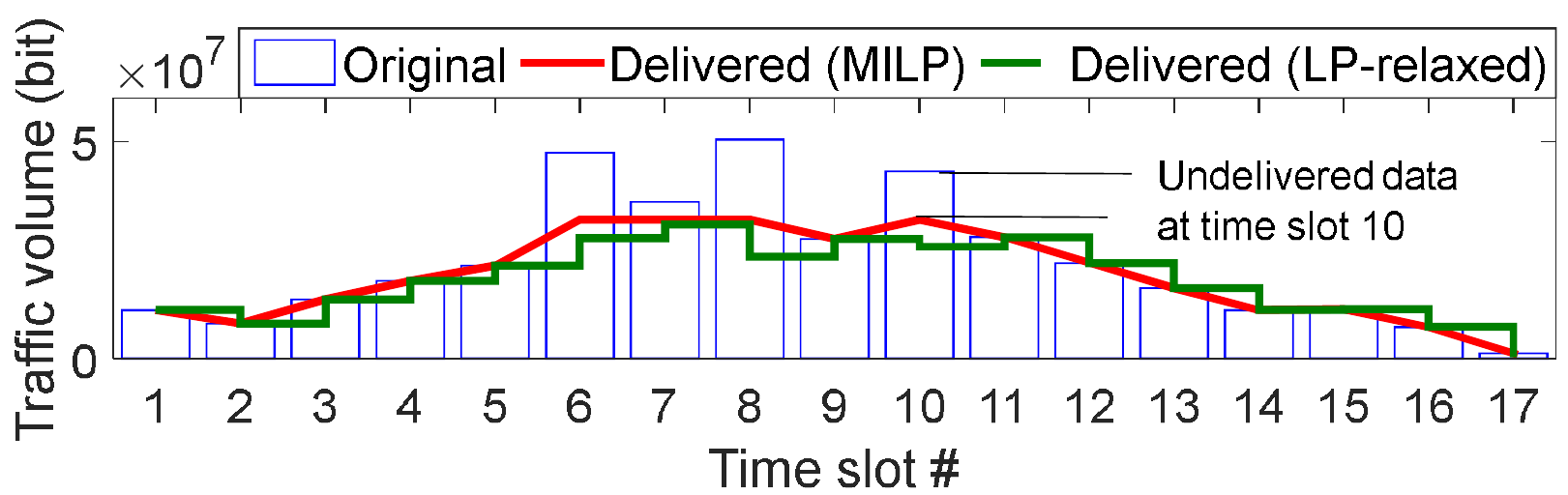}
\caption{\textbf{Overall delivered quality layers in each time slot}}
\label{code192021}
\end{figure}
\begin{figure}[t]
\includegraphics[width=\linewidth]{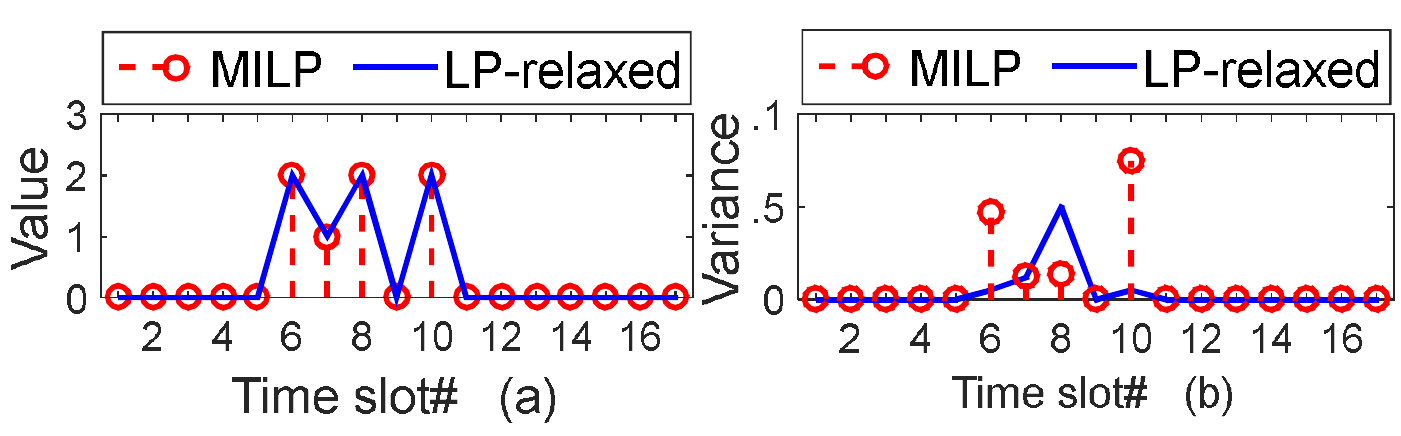}
\caption{\textbf{Comparing fairness of MILP and LP-relaxed models}}
\label{code18}
\end{figure}

In the next step, we present the performance evaluation of the proposed MILP and LP-relaxed models with respect to different metrics. At first, we compare the execution time of the models written in Python and Pulp library \cite{pulp}. Employing the default and extended network topology and increasing the switch links bandwidth from 8 to 100 Mbps, we run experiments with 10, 20, 30,  and 40 clients, with the results showing the practical applicability of the algorithm (refer to Table \ref{T4}). As seen in Table \ref{T4}, our MILP model has a significantly higher execution time. With the growth of network size, this difference can be even higher. On the other hand, \textcolor{black}{ the results show that for limited number of clients and switches, the relaxed model is more applicable than the MILP} and can be used in bigger networks. We extend the investigation by comparing the measured values of objective functions in both models according to the default topology. As presented in Fig. \ref{code131516}, the obtained objective values of our LP-relaxed model are lower than those of MILP in some time slots (e.g. 6, 7, 8, and 10). The average of delivered quality layer ($T_c$) and quality switches ($I_c$) in each time slot are given in Fig. \ref{code131516}. As we observe, the average of the delivered quality layer and also the quality switches obtained by the LP-relaxed model report a lower performance in comparison with the MILP model in some time slots. The first possible reason for this is the rounding down of variable $\chi_c$ in Eq.\ref{eq16}. Another explanation can be the calculation of $\bar{\delta}_c$ as the average segment size of the video requested by client $c$. Moreover, with the default parameters, the values of the delivered quality layer obtained by the LP-relaxed and MILP models are shown in Fig. \ref{figg1555} (a) and (b), respectively. Although the MILP and LP-relaxed models have different treatments in delivering quality layers, especially in time slots $6, 7, 8 ,$ and $10$, the amount of sent data in these time slots are almost equal (See Fig. \ref{code192021}). In the final experiment, we compare the maximum received quality to its maximum supported value $m_c$ for the MILP and LP-relaxed models. We show these parameters as $max\{m_c-\sum_l\omega_{cl}\mid\forall c\}$ and $max\{m_c-\chi_c\mid\forall c\} $, respectively (see Fig. \ref{code18} (a)). Although Fig. \ref{code18} (a) shows identical behavior, the LP-Relax model surpasses MILP in the variance of the delivered quality layer (see Fig. \ref{code18} (b)). Note that the variances are measured based on the normalized values of the delivered quality layers.

\section{Conclusion and Future Work}
Today, dynamic adaptive streaming over HTTP (DASH) is emerging the distinguished technology for delivering video over the Internet; it is indebted to the simplicity and the efficiency of HTTP protocol. DASH enables clients to adjust their adaptation according to observed feedback from the network. Moreover, DASH provides an opportunity to serve clients by media cache servers at the edge of the network. Delegating an optimal adaptation process to clients has suffered as an issue where adaptation is performed based on local information on the client. Furthermore, achieving maximum QoE metrics and QoE-fairness cannot be guaranteed by employing local parameters for the clients. 

By leveraging the SDN paradigm, we proposed a new SDN-based framework, named $S^2VC$, to address the problem of rate adaptation by focusing on maximizing QoE and QoE-fairness. We elaborated on the architecture of $S^2VC$ by designing a set of interconnected components. After collecting critical data from the network, such as client requests and network resources, $S^2VC$ determined an appropriate data path and rates in a time slot based manner. In fact, in $S^2VC$, upon the receipt of a request packet from the DASH client, it was sent to the SDN controller as a Packet-In. After processing the Packet-In by the proposed application modules, an appropriate solution was achieved. Then, the SDN controller configured the OpenFlow switches and HTTP-media server to start packet transmission.

We formulated the problem as a MILP optimization model and showed that it is an NP-complete problem. We further extended our approach by proposing an LP-relaxation model to provide practical applicability for the proposed framework. Regarding the performance evaluation, we expanded the Scootplayer and implemented the proposed framework and its components in Python. Then, by employing Mininet and Floodlight, we conducted experiments in different scenarios, evaluated the QoE-fairness and QoE metrics, and made comparison with different state-of-the-art network-based approaches. The results validated the performance of the proposed framework.

We have two main directions as the  future work. The first one is to extend our proposed framework for live multicast streaming in the network edge. Also, we can evaluate the usage of HTTP2.0 and Google Quick UDP Internet Connections (QUIC) protocol in the context of DASH-based services.

\bibliographystyle{unsrt}
\bibliography{main.bib}

\end{document}